\documentclass[iop,apj,twocolumn,numberedappendix]{openjournal}
\usepackage{graphicx}
\usepackage{xcolor}
\usepackage{newtxtext,newtxmath,enumitem,multirow,ctable}
\usepackage{bm}
\usepackage{hyperref}
\hypersetup{breaklinks,colorlinks,citecolor=blue,urlcolor=cyan}

\newcommand{\orcidauthor}[3]{\author{#2$^{#3}$ \href{http://orcid.org/#1}{\includegraphics[scale=0.25]{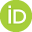}}}}
\newcommand{\Msun}{\,{\rm M}_\odot}
\newcommand{\feh}{{\rm [Fe/H]}}

\newcommand{\Mgc}{M_{{\rm GC}}}
\newcommand{\Ngc}{N_{{\rm GC}}}

\newcommand{\Mh}{M_{{\rm h}}}

\newcommand{\tlb}{t_{\rm lookback}}
\newcommand{\curse}{\textit{curse of dimensionality}}
\newcommand{\Jr}{J_r}
\newcommand{\logJr}{\log J_r}
\newcommand{\Jp}{J_\phi}
\newcommand{\Jz}{J_z}
\newcommand{\logJz}{\log J_z}
\newcommand{\logr}{\log r}
\newcommand{\rperi}{r_{\rm peri}}
\newcommand{\rapo}{r_{\rm apo}}
\newcommand{\logrperi}{\log r_{\rm peri}}
\newcommand{\lograpo}{\log r_{\rm apo}}
\newcommand{\E}{E}

\newcommand{\age}{{\rm age}}

\defcitealias{choksi_formation_2018}{CGL18}
\defcitealias{chen_modeling_2022}{CG22}
\defcitealias{chen_formation_2023}{CG23}
\defcitealias{chen_catalogue_2024}{CG24}
\defcitealias{massari_origin_2019}{M19}
\defcitealias{belokurov_-situ_2024}{BK24}

\shorttitle{Galaxy assembly revealed by globular clusters}
\shortauthors{Y. Chen \& O. Y. Gnedin}

\begin{document}

\title{\vspace{-6mm}Galaxy assembly revealed by globular clusters\vspace{-14mm}}
\orcidauthor{0000-0002-5970-2563}{Yingtian Chen}{\star}
\orcidauthor{0000-0001-9852-9954}{Oleg Y. Gnedin}{}
\affiliation{Department of Astronomy, University of Michigan, Ann Arbor, MI 48109, USA}
\thanks{$^\star$E-mail: \href{mailto:ybchen@umich.edu}{ybchen@umich.edu}}

\begin{abstract}
Many observable properties of globular clusters (GCs) provide valuable insights for unveiling the hierarchical assembly of their host galaxy. For the Milky Way (MW) in particular, GCs from different accreted satellite galaxies show distinct chemical, spatial, kinematic, and age distributions. Here we examine such clustering features for model GC populations in simulated galaxies, which are carefully selected to match various observational constraints of the MW assembly. We evaluate several widely used clustering, dimensionality reduction, and supervised classification methods on these model GCs, using 10 properties that are observable in the MW. We can categorize \textit{in-situ} and \textit{ex-situ} formed GCs with about 90\% accuracy, based solely on their clustering features in these 10 variables. The methods are also effective in distinguishing the last major merger in MW-analogs with similar accuracy. Although challenging, we still find it possible to identify one, and only one, additional smaller satellite. We develop a new technique to classify the progenitors of MW GCs by combining several methods and weighting them by the validated accuracy. According to this technique, about 60\% of GCs belong to the \textit{in-situ} group, 20\% are associated with the Gaia-Sausage/Enceladus event, and 10\% are associated with the Sagittarius dwarf galaxy. The remaining 10\% of GCs cannot be reliably associated with any single accretion event.
\end{abstract}


\maketitle

\section{Introduction}
\label{sec:intro}

Recent observations have significantly deepened our understanding of the hierarchical assembly process of the Milky Way (MW) galaxy. Pioneering analysis of the stellar abundances and kinematics in the outer regions of the Galactic bulge by \citet{ibata_dwarf_1994} was the first to uncover evidence of a merging event: the Sagittarius dwarf galaxy. Subsequent kinematic studies of metal-poor stars in the solar neighbourhood \citep{helmi_debris_1999} identified debris streams of another merger. These streams are now commonly referred to as the Helmi streams.
More recently, \citet{deason_progenitors_2015} proposed potential accretion of massive satellite galaxies based on an analysis of the ratio of blue straggler stars to blue horizontal branch stars in the Galactic halo. 

The \textit{Gaia} mission \citep{gaia_collaboration_gaia_2016,gaia_collaboration_gaia_2018,gaia_collaboration_gaia_2023} has enabled comprehensive investigation of more sophisticated structures of the Galaxy in the 6-dimensional phase space \citep[see a review by][]{deason_galactic_2024}. Using \textit{Gaia} data, \citet{belokurov_co-formation_2018}, \citet{helmi_merger_2018}, and \citet{deason_apocenter_2018} confirmed the merger with a massive satellite through chemical and kinematic analyzes of both disk and halo stars. In this work we refer to this satellite as the Gaia-Sausage/Enceladus (GS/E). Additionally, the \textit{Gaia} data played an important role in identifying or substantiating smaller-scale merger events. These include the Sequoia event \citep{myeong_evidence_2019}, Thamnos structure \citep{koppelman_characterization_2019}, Cetus Stream \citep[][initially proposed by \citealp{newberg_discovery_2009}]{yuan_revealing_2019}, LMS-1/Wukong Stream \citep{yuan_low-mass_2020,naidu_evidence_2020,malhan_evidence_2021}, Pontus \citep{malhan_global_2022}, and Antaeus \citep{oria_antaeus_2022,ceccarelli_walk_2024}.

All of these studies share the same assumption: stellar populations from past mergers retain their original properties that set them apart from both the \textit{in-situ} population and from other mergers. These properties include chemical, spatial, kinematic, and age information, and are inherited from their progenitor galaxies. Therefore, to fully uncover the MW assembly history, it is critical to compile a comprehensive sample of stellar tracers. This includes single stars, stellar streams, and globular clusters (GCs). However, current samples of single stars and stellar streams in the MW are incomplete. As a result, one often relies on a subset of these objects, selected preferentially based on existing knowledge of known mergers. This approach leads to inefficiency and bias in the detection of previously unknown mergers \citep{malhan_global_2022}.

GCs are compact objects with ages typically exceeding 10~Gyr. They serve as ideal tracers of ancient mergers, which have largely dispersed their stellar components into the stellar halos of central galaxies. The sample of MW GCs is relatively more complete than that of single stars and streams, thanks to their compactness and high luminosity. This motivates a growing amount of research \citep[e.g.,][]{massari_origin_2019,forbes_reverse_2020,callingham_chemo-dynamical_2022,malhan_global_2022,belokurov_-situ_2024} employing GCs to investigate the assembly history of the MW. These studies have utilized diverse classification algorithms on different subsets of GC characteristics, such as the age--metallicity space, integral of motion (IoM) space, or orbital action space. They also relied on previously established knowledge of galactic mergers, which vary across studies. Consequently, while these studies have reached a general agreement on distinguishing \textit{in-situ} from \textit{ex-situ} clusters, the assignment of \textit{ex-situ} GCs to specific progenitors remains debated except for the most significant mergers. This uncertainty arises from the reliance on different combinations of GC characteristics, classification methods, and prior knowledge.

To validate these classifications requires knowing the correct label for each GC, which is impossible for MW GCs due to the limitation of having only present-day observations. Therefore, we must turn to galaxy formation simulations, where the progenitor of every simulation particle is precisely known. Unfortunately, directly modeling star cluster formation within such simulations remains challenging. The efficiency of cluster formation is regulated by the complicated and entangled feedback processes \citep[see, e.g.,][]{li_star_2018,grudic_starforge_2021,brown_testing_2022}. Running full cosmological simulations of realistic galactic environments to the present day is extremely computationally expensive. If the simulations do not proceed to the present day, their subgrid parameters remain poorly constrained as we cannot compare the properties of model GCs with the observations. As a result, these simulations may not accurately match the Galactic GC system.

Fortunately, there is an alternative approach to modeling GC systems by post-processing existing simulation suites with analytical GC formation and evolution models. These models not only match observed GC properties more closely by calibrating model parameters but also generate substantially larger samples of model GCs in a variety of galactic environments. Thus, post-processing models are an ideal tool for creating mock catalogs of GCs, which are crucial for the evaluation of classification algorithms. Existing post-processing models include the MOdelling Star cluster population Assembly In Cosmological Simulations (MOSAICS) model for cluster formation and evolution \citep{kruijssen_photometric_2008,kruijssen_evolution_2009,kruijssen_modelling_2011}, and the models developed by \citet{renaud_origin_2017}, \citet{creasey_globular_2019}, \citet{phipps_first_2020}, \citet{halbesma_globular_2020}, and \citet{valenzuela_globular_2021}.

\citet{pfeffer_e-mosaics_2018} and \citet{kruijssen_e-mosaics_2019} conducted the E-MOSAICS project by implementing the MOSAICS model into the EAGLE simulations \citep{schaye_eagle_2015,crain_eagle_2015}. This project matched various observational properties of young star clusters. Using the E-MOSAICS data, subsequent studies by \citet{kruijssen_e-mosaics_2019}, \citet{pfeffer_predicting_2020}, \citet{reina-campos_mass_2020}, and \citet{carlsten_structures_2021} studied the relationship between GC properties and the assembly history of their host galaxies. More recently, \citet{trujillo-gomez_situ_2023} developed a supervised classification framework based on the E-MOSAICS data. This framework is capable of distinguishing \textit{in-situ} from \textit{ex-situ} clusters with accuracy up to 80--90\%.

Another example of post-processing models is the earlier versions of our model \citep{muratov_modeling_2010,li_modeling_2014,choksi_formation_2018,choksi_formation_2019,choksi_origins_2019}. These versions take the halo merger tree as the only input and trigger a cluster formation event upon detecting a major merger or fast accretion. A sequence of scaling relations is then employed to analytically calculate the mass and metallicity of individual clusters. Additionally, the model accounts for the cluster mass loss due to stellar evolution and tidal disruption during the subsequent evolution. We extended the model starting with \citet[][hereafter \citetalias{chen_modeling_2022}]{chen_modeling_2022} by introducing the tagging technique to assign collisionless simulation particles as GC tracer particles. This provides the spatial and kinematic information about GC systems at present. Moreover, by tracking the tidal field along the trajectories of these GC particles, our model now allows for more precise calculations of tidal disruption. Our model successfully aligns with various observed scaling relations of GC systems. These include the GC effective radius--halo mass relation and the GC mass/number--halo mass relation down to a halo mass $\Mh\sim10^8\Msun$ \citep{chen_formation_2023}. 
 
Furthermore, by applying our model to simulated galaxies carefully selected to match the assembly history of the MW, we have generated a catalog of model GCs in MW-analog systems \citep[][hereafter \citetalias{chen_catalogue_2024}]{chen_catalogue_2024}. This catalog provides key observables such as the mass, metallicity, orbital actions, circularity, galactocentric radius, pericenter/apocenter radii, specific energy, and age. These variables cover most of the properties traditionally utilized to trace the progenitors of GCs. This catalog accurately reproduces important characteristics of the Galactic GCs, including their mass function, metallicity distribution, radial profile, and total velocity dispersion.

Using two Illustris TNG50-1 \citep[][hereafter TNG50]{nelson_first_2019,pillepich_first_2019,nelson_illustristng_2021} galaxies from the \citetalias{chen_catalogue_2024} catalog and two galaxies from the Latte suite of the FIRE-2 simulations \citep{wetzel_reconciling_2016}, our goal in this paper is to find the most informative properties for revealing the host galaxy assembly. Additionally, we aim to evaluate the optimal algorithm for identifying the progenitor galaxies of GCs. We begin with splitting the model GCs into \textit{in-situ} and \textit{ex-situ} groups. We then concentrate on the \textit{ex-situ} clusters to differentiate between GCs from different progenitor galaxies. We examine five widely used unsupervised clustering methods, which are then enhanced by dimensionality reduction and supervised classification techniques. After determining the most relevant properties and the most effective algorithm, we repeat the same process with MW GCs. This allows us to identify the most significant assembly features of the Galaxy.

This paper is structured as follows. In \S\ref{sec:model_sample} we provide a brief overview of the GC formation model and introduce the sample of model GC systems analyzed in this study. Next, in \S\ref{sec:in_vs_ex} we explore various classification methods to distinguish \textit{in-situ} and \textit{ex-situ} clusters. In \S\ref{sec:progenitors} we identify individual merger events, focusing on the \textit{ex-situ} clusters. In \S\ref{sec:mw} we apply the methods from previous sections to classify the MW GCs. Our discussion in \S\ref{sec:discussion} investigates the extent to which \textit{ex-situ} clusters preserve the kinematics of their progenitor hosts (\S\ref{sec:evolution}), and how the GC properties from a single merger can be used to reconstruct significant characteristics of that progenitor (\S\ref{sec:merger_properties}). We also compare this work with a related study in \S\ref{sec:T23}. In \S\ref{sec:summary} we summarize our key findings. We provide a rigorous definition of classification accuracy employed throughout this work in Appendix~\ref{sec:accuracy}, where we also calculate the minimum achievable accuracy. We discuss the maximum achievable accuracy in Appendix~\ref{sec:max_accuracy}.

\section{Model sample}
\label{sec:model_sample}

In this work, we analyze GC systems within four simulated galaxies. Two of them are taken from the TNG50 simulation, while the other two are from the Latte suite of the \mbox{FIRE-2} simulations. The GC systems are generated using the \citetalias{chen_catalogue_2024} version of our model. This section provides a brief overview of the model setup, the background simulations, and the specific galaxy samples selected for our analysis.

\begin{figure*}
    \centering
    \includegraphics[width=\linewidth]{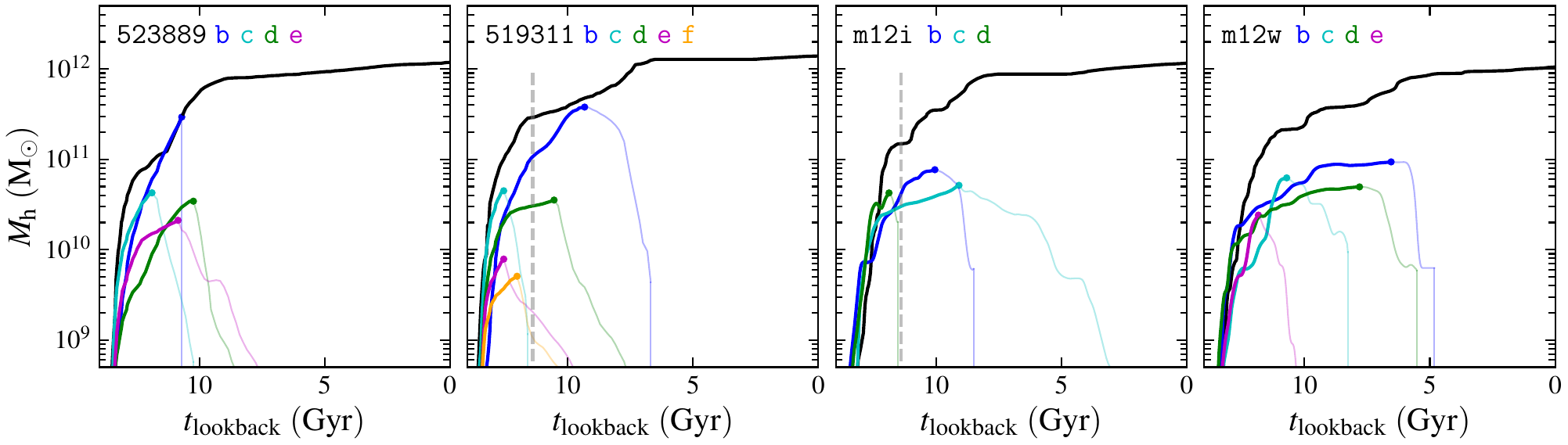}
    \vspace{-3mm}
    \caption{Mass assembly histories of the four sample galaxies (black lines). The colored lines in each panel stand for the major satellites that bring at least 3 GCs to the central galaxy, with solid circles marking the maximum masses. The lines are smoothed for visual clarity. The dashed vertical lines at $\tlb=11.4$~Gyr stand for the threshold of \textit{early mergers}. We keep the same color and naming scheme throughout the paper when referring to these satellites.}
    \label{fig:mh_log_vs_tlookback}
\end{figure*}

\subsection{Recap of the model}
\label{sec:model}

We do not repeat the full details of our model as these have been extensively described and examined for galaxies of a wide range of mass and assembly history in our prior works. Instead, we briefly summarize the four key steps and introduce the three free parameters of the model.
\begin{enumerate}[labelindent=0pt,labelwidth=10pt,labelsep*=0pt,leftmargin=!,align=parleft]
    \item \textbf{Cluster formation} is triggered by rapid mass growth of the host galaxy, quantified by the specific mass accretion rate $\dot{M}_{\rm h}/\Mh>p_3$. The threshold $p_3$ is a model parameter controlling the cluster formation frequency, with a typical value $\sim1\ {\rm Gyr^{-1}}$. When a cluster formation event is triggered, we compute the total mass of newly formed clusters using a sequence of scaling relations, including the stellar mass--halo mass relation \citep{behroozi_average_2013}, the gas mass--stellar mass relation \citep{lilly_gas_2013,genzel_combined_2015,tacconi_phibss_2018,wang_3_2022}, and the linear gas mass--cluster mass relation \citep{kravtsov_formation_2005}, $M_{\rm tot}=1.8\times10^{-4}p_2M_{\rm gas}$. Here $p_2$ is another model parameter controlling the cluster formation rate, with a typical value $\sim 10$ for clusters with initial mass $>10^4\Msun$. We also calculate cluster metallicity in this step using a galaxy mass--metallicity relation calibrated to recent data from the \textit{James Webb Space Telescope} \citepalias[see Appendix~A in][]{chen_catalogue_2024}. We use a combination of Gaussian process and Gaussian noise to model the scatter in these relations.
    \item \textbf{Cluster sampling} determines the initial mass of individual clusters. We sample the initial mass via a Schechter function with an exponential cutoff at $10^7\Msun$. We only consider clusters with initial mass $>10^4\Msun$ because less massive clusters are likely to be tidally disrupted in less than 1~Gyr.
    \item \textbf{Particle assignment} selects collisionless simulation particles as ``tracer'' particles of individual clusters. We first consider newly formed (age $<10$~Myr) star particles within 3~kpc (approximately twice the effective radius of newly formed stars at the epochs when GC formation is most active) from the galactic center. If there are not enough such particles, we turn to older star particles or dark matter particles near the galactic center. We only use these particles to trace the positions and velocities of GCs, and calculate all other physical properties analytically.
    \item \textbf{Cluster evolution} takes into account two effects: stellar evolution and tidal disruption. Stellar evolution is modeled as an instantaneous 45\% mass loss at formation following \citet{gieles_mass-loss_2023}. Such an approximation is appropriate since the stellar evolution time scale is  much shorter than the typical lifetime of GCs ($\gtrsim 10$~Gyr). We model the tidal disruption rate as a power-law function with the initial cluster mass, current mass, and the strength of tidal field based on the eigenvalues of the tidal tensor. We compute the tidal tensor numerically using the scheme in \citetalias{chen_modeling_2022}. This scheme evaluates the gravitational potential on a $3\times3\times3$ cubic grid and computes the second order finite difference to approximate the tidal tensor. We correct the systematic bias of the approximation by multiplying the numerically-derived tidal tensor by the third model parameter $\kappa\gtrsim1$, which controls the intensity of tidal disruption.
\end{enumerate}
The numerical implementation of the model is publicly available at \url{https://github.com/ybillchen/GC_formation_model} under the BSD 3-Clause License, which allows redistribution and modification of the source code with moderate limitations.

\subsection{Illustris TNG50 simulations}

The model galaxies in the TNG50 simulation are detailed in \citetalias{chen_catalogue_2024}. We have calibrated the best model parameters to be $(p_2,p_3,\kappa)=(14,0.5\,{\rm Gyr^{-1}},1.5)$. The two selected galaxies, \texttt{523889} and \texttt{519311}\footnote{The galaxy IDs are \texttt{SubfindID} in the TNG50 \textsc{subfind} catalog.}, are specifically chosen for their similarity to the Milky Way in terms of virial mass, circular velocity, and key assembly characteristics. We repeat the model on these two galaxies with different random seeds to select the realizations that most accurately reflect the observed total mass, mass function, metallicity distribution, radial profile, and velocity dispersion of the MW GC system. 

In Fig.~\ref{fig:mh_log_vs_tlookback}, we present the mass assembly histories of the galaxies \texttt{523889} and \texttt{519311}. These galaxies experienced major mergers similar to the GS/E at $\tlb\approx8-10$~Gyr, followed by a period of quiescent mass growth. Additionally, we plot the mass histories of important satellite galaxies that contributed at least three GCs surviving to the present day. For convenience, we adopt a naming convention similar to that commonly used for planetary systems: the central galaxy ID followed by a lowercase letter starting with \texttt{b}. For example, \texttt{523889-b} denotes the most massive satellite of \texttt{523889}.

\begin{figure*}
    \centering
    \includegraphics[width=0.9\linewidth]{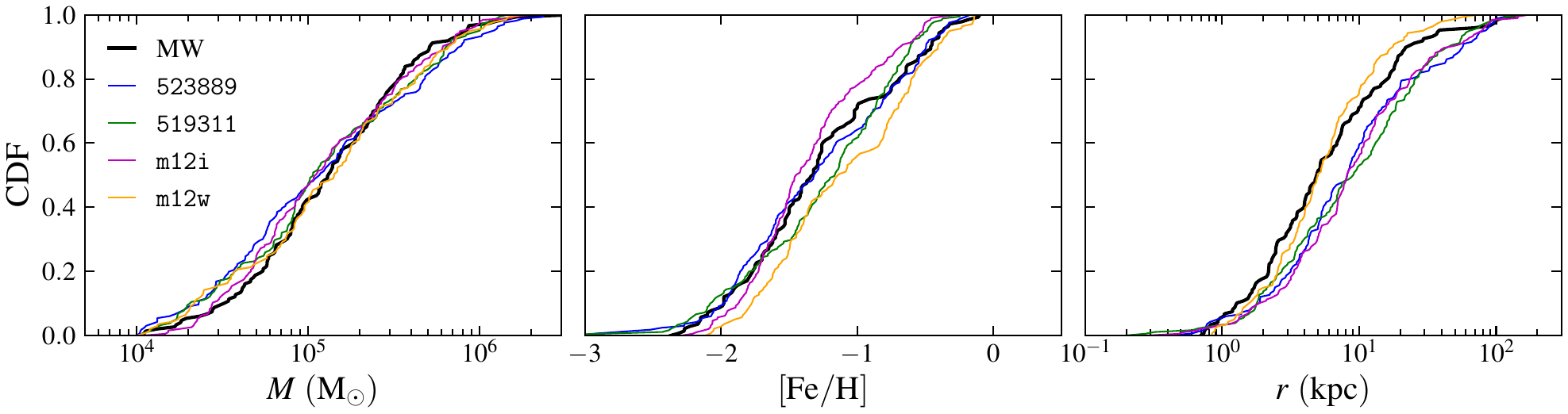}
    \vspace{0mm}
    \caption{Cumulative distribution functions of cluster mass (\textit{left}), metallicity (\textit{middle}), and radius (\textit{right}). We show the observed value of the MW GCs as black lines. Only clusters with $M>10^4\Msun$ are taken into account.}
    \label{fig:cdf}
\end{figure*}

\subsection{FIRE-2 simulations}

In addition to the TNG50 simulations, we apply our model to the Latte simulation suite of the FIRE-2 public data release\footnote{Available at \url{https://flathub.flatironinstitute.org/fire}.} \citep{wetzel_public_2023}. The FIRE-2 cosmological zoom-in simulations are part of the Feedback In Realistic Environments (FIRE) project, generated using the \textsc{gizmo} code \citep{hopkins_new_2015} and the FIRE-2 physics model \citep{hopkins_fire-2_2018}. 

The Latte suite is a set of zoom-in simulations of isolated galaxies similar to the MW. The mass resolution is $3.5\times10^4\Msun$ for DM particles and $7.07\times10^3\Msun$ for stellar and gas particles initially. Such high mass resolution enables the simulation to effectively resolve structures within a scale of $\approx25$~pc in the central galaxy \citep{wetzel_reconciling_2016}. 

The Latte suite uses the following cosmological parameters: $\Omega_{\rm m}=0.272$, $\Omega_{\rm b}=0.0455$, $\Omega_\Lambda=0.728$, $h=0.702$, $\sigma_8=0.807$, and $n_s=0.961$. Each simulation outputs 601 snapshots from $z=99$ to the present day, with a typical time interval $\approx20$~Myr.

The halo catalogs are generated using the \textsc{rockstar} halo finder \citep{behroozi_rockstar_2013} on dark matter particles. As a result, the halo mass provided in these catalogs represents only the dark matter mass. To obtain the total halo mass, we multiply this value by the correction factor $(1-f_{\rm b})^{-1}$, where $f_{\rm b}\equiv\Omega_{\rm b}/\Omega_{\rm m}$ is the cosmological baryon fraction. Based on the halo catalogs, the merger trees are generated with the \textsc{consistent tree} code \citep{behroozi_gravitationally_2013}. 

The publicly available data include the complete merger tree and halo catalogs across 601 output epochs, along with a subset of 39 full particle snapshots that begin at $z=10$. The asynchronous output frequencies between the merger tree and the particle snapshots require specific adaptations to our model. These modifications are outlined in the next subsection.

We select two of the simulated Latte galaxies, \texttt{m12i} and \texttt{m12w}. These galaxies resemble the MW with a virial mass $\approx10^{12}\Msun$. For comparison with the TNG50 galaxies, we also plot the mass assembly histories of the FIRE-2 galaxies and their satellite galaxies in Fig.~\ref{fig:mh_log_vs_tlookback}. Notably, the FIRE-2 galaxies reproduce the key characteristics of the MW. They also accumulated the majority of their mass early, with a GS/E-like merger at $\tlb=5-10$~Gyr.

\subsubsection{Model adaptions to FIRE-2}

Given that the cluster formation step depends solely on the merger tree, while the particle assignment step requires both the merger tree and particle outputs, we must execute these two steps asynchronously and synchronize the outcomes at ``full'' snapshots, where both the merger tree and particle outputs are available. 

If a cluster formed at a full snapshot, we employ the same methodology as in \citetalias{chen_modeling_2022} to select GC tracer particles. Otherwise, we track the merger tree to locate the host galaxy's descendant in the next full snapshot and then choose collisionless particles from this snapshot to represent the cluster. Similarly to the approach in \citetalias{chen_modeling_2022}, we select stellar particles within 3~kpc from the galactic center, formed between $t_{\rm form}$ and $t_{\rm form}+\Delta t$, where $t_{\rm form}$ is the cluster's formation lookback time (not the full snapshot's lookback time), and $\Delta t$ is initially set to 10~Myr. If this criterion yields insufficient number of stellar particles, we expand $\Delta t$ to the time interval between the adjacent snapshots. However, this method is not always efficient due to the typical time interval of the FIRE-2 simulation being only 20~Myr. In the rare cases ($\lesssim10\%$) where there are still not enough stellar particles, we use dark matter particles near the galactic center as GC tracers.

We test the effectiveness of this synchronization approach by adding 20 more snapshots at $z\gtrsim1$ for the \texttt{m12i} simulation in addition to the original 39 publicly available full snapshots. This shortens the time interval during the peak period of GC formation from up to 1~Gyr to around 200~Myr. The latter is similar to that in TNG50. However, the resulting GC system show no systematic difference to the original 39 outputs, suggesting that such an approach is robust to different numbers of synchronization points. Therefore, we use only the original 39 public snapshots for the other Latte galaxy, \texttt{m12w}.

In addition, we improve the calculation of tidal tensor for the Latte suite. As suggested by \citetalias{chen_modeling_2022} (see \S2.2.3 and Appendix~A therein), the optimal grid size for numerically computing the tidal tensor should be $\gtrsim3$ times larger than the typical spatial resolution where most GCs reside. Accordingly, we choose a grid size of 100~pc for the Latte suite, reduced from the 300~pc used in the TNG50 simulations. This adjustment takes advantage of the finer spatial resolution in the Latte suite $\approx 25$~pc.

We emphasize that the optimal model parameters for the Latte suite simulations are not necessarily the same as those used for TNG50 because of the distinct output frequencies of the two simulation suites. Therefore we re-calibrate the model following the same process as in \citetalias{chen_catalogue_2024}. After that, we repeat the model 128 times to select the realization that most closely match the observational constraints of the MW GC system.

The optimal parameters for the Latte suite are $(p_2,p_3,\kappa)=(7,1\,{\rm Gyr^{-1}},1.5)$. Compared to the best values for TNG50, the $p_2$ value is lower due to the higher output frequency of the Latte simulations, by a factor of six. Moreover, the more frequent outputs not only resolve more galaxy assembly events but also record tiny and sometimes spurious mass changes of the host galaxy. In other words, both ``signal'' and ``noise'' are magnified. Consequently, we need a higher trigger threshold $p_3$ to filter out the noise. We publicly release the two best MW-analogs from FIRE-2 at \url{https://github.com/ognedin/gc_model_mw}, where we have also released the two TNG analogs from \citetalias{chen_catalogue_2024}.

With such parameters, we show the cumulative cluster mass function, metallicity distribution, and radial profile of the five simulated galaxies in Fig.~\ref{fig:cdf}. Most distributions can match observed distributions of the MW GCs, indicating that these simulated GC systems are close analogs to the MW system.

\begin{figure*}
    \centering
    \includegraphics[width=\linewidth]{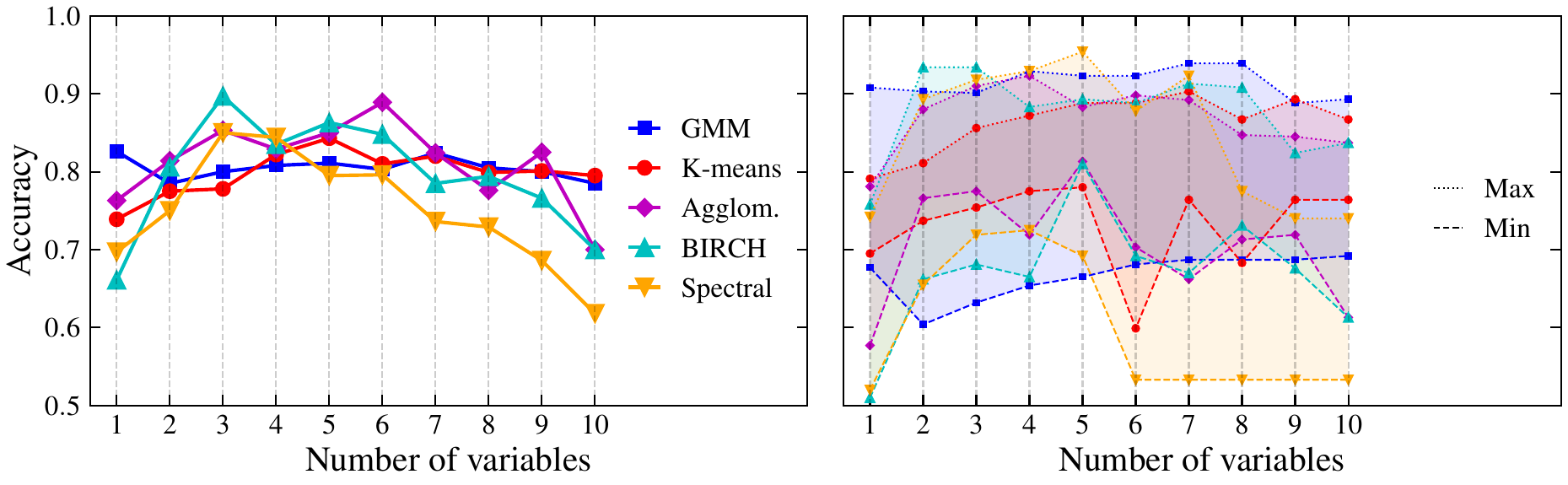}
    \vspace{-4mm}
    \caption{Median (\textit{left}) and minimum/maximum (\textit{right}) classification accuracy of the four sample galaxies against the number of observable variables. For each method, we first identify the best configuration of variables producing the highest median accuracy over the four galaxies. Starting from that configuration, we sequentially remove/add one variable each time to obtain the configurations producing the highest accuracy with fewer/more number of variables.}
    \label{fig:dim_acc_no_threshold}
\end{figure*}

\section{Classification of \textit{in-situ} vs. \textit{ex-situ} clusters}
\label{sec:in_vs_ex}

In this section, we employ classification methods to distinguish \textit{in-situ} from \textit{ex-situ} formed clusters, taking into account a range of observable properties. Our analysis incorporates ten properties spanning the chemical, spatial, kinematic, and age aspects: metallicity $\feh$, orbital actions $\Jr$, $\Jp$\footnote{In our case, $\Jp=L_z$, the specific angular momentum along $z$-axis.}, and $\Jz$, circularity $\varepsilon$, galactocentric radius $r$, pericenter/apocenter radii $\rperi$ and $\rapo$, specific energy $\E$, and age. For properties spanning many orders of magnitudes we use the logarithmic values: $\logJr$, $\logJz$, $\logr$, $\logrperi$, $\lograpo$. For visual clarity, we omit the subscript ``10'' in 10-based logarithms and use ``$\ln$'' for natural logarithms throughout the paper. A detailed description of how these properties are computed can be found in \citetalias{chen_catalogue_2024}.

We define the clusters that formed within the central galaxy as \textit{in-situ}, whereas those originated from satellites and accreted through mergers are referred to as \textit{ex-situ}. However, this distinction is less clear during the early stages of galaxy assembly. During this period, star clusters formed in massive ``satellites'', such as \texttt{519311-c} and \texttt{m12i-d}, might have properties similar to \textit{in-situ} clusters, as these satellites were potentially more massive than the central galaxy at the time of their GC formation, see Fig.~\ref{fig:mh_log_vs_tlookback}. Consequently, differentiating such GCs accreted early from the strictly \textit{in-situ} population may not be meaningful.

We observe that these early satellites typically reached their maximum mass more than 12~Gyr ago and were mostly dissolved at $\tlb\gtrsim 11.4$~Gyr. Based on this, we refer to the clusters accreted more than 11.4~Gyr ago as \textit{early-accreted}. In the following analysis, we generally combine the \textit{in-situ} and \textit{early-accreted} populations, except in \S\ref{sec:evolution} and \S\ref{sec:merger_properties} where we investigate the properties of individual mergers.

In the remainder of this section, we evaluate three classification procedures: 1) unsupervised clustering methods, 2) combining dimensionality reduction with unsupervised clustering methods, and 3) combining unsupervised clustering methods with supervised classification methods. We conduct the evaluation by comparing each classification with the true labels provided by the model. The primary goal is to identify the most effective observables and classification methods that can accurately identify the \textit{in-situ} and \textit{ex-situ} origins of GCs.

\subsection{Direct clustering}
\label{sec:direct_clustering}

We have tested five widely used unsupervised clustering methods. These algorithms are implemented through the \textsc{scikit-learn} package \citep{pedregosa_scikit-learn_2011}\footnote{Detailed descriptions of all methods can be found at \url{https://scikit-learn.org/stable/modules/clustering}.}. 
\begin{itemize}[labelindent=0pt,labelwidth=10pt,labelsep*=0pt,leftmargin=!,align=parleft]
    \item \textbf{K-means} categorizes data points into K distinct groups\footnote{We use the term ``group'' instead of the commonly used ``cluster'' to avoid confusion with ``star cluster''.}. This method minimizes the dispersion of data points within each group. Finding the global optimal classification is computationally challenging. Therefore, we employ the \textsc{scikit-learn} implementation of K-means, which uses an iterative algorithm to find local minima. The initial setup for this algorithm is determined using the K-means++ method \citep{arthur_k-means_2007}, which helps select effective starting state for the clustering process.
    \item \textbf{Gaussian Mixture Model (GMM)} assumes that the data points are drawn from a finite number of Gaussian distributions. The implementation of GMM in \textsc{scikit-learn} utilizes a expectation-maximization algorithm to accurately fit multi-dimensional data. The likelihood of a GMM with $N_{\rm g}$ components is defined as:
    \begin{equation}
        {\cal L} = \prod_{i=1}^n\sum_{j=1}^{N_{\rm g}} \tau_{ij}\ln {\cal N}(\bm{x}_i|\bm{\mu}_j,{\bf\Sigma}_j).
        \label{eq:gmm}
    \end{equation}
    Here, $\bm{x}_i$ represents the $i$-th data point, while ${\cal N}$ denotes the Gaussian function, characterized by its mean vector $\bm{\mu}_j$ and covariance matrix ${\bf\Sigma}_j$, both in $N$ dimensions. The term $\tau_{ij}$ denotes the probability of the $i$-th data point belonging to the $j$-th Gaussian component. In our implementation, each component has its own covariance matrix. We initialize the GMM with K-means clustering and iterate up to 100 times until the model converges.
    \item \textbf{Agglomerative Clustering} constructs a hierarchy of groups in a bottom-up approach. Initially, each data point is considered as an individual group. These groups then merge to form larger groups recursively until there is a single group containing all data points. The \textsc{scikit-learn} package offers four merging strategies. \texttt{Ward} minimizes the variance within the merging groups. The \texttt{average}, \texttt{maximum}, and \texttt{single} strategies merge groups based on the lowest average, maximum, or minimum distance between members of the groups, respectively. We find that \texttt{Ward} performs the best among the four strategies.
    \item \textbf{Balanced Iterative Reducing and Clustering using Hierarchies (BIRCH)} is a hierarchical clustering method in a top-down manner. It begins by constructing a clustering feature (CF) tree from the input data, where each data point is treated as a multi-dimensional vector. The CF for a set has three attributes: the number of elements in the set, the linear sum of all vectors, and the square sum of all vectors. A CF tree node has several sub-clusters, each associated with a CF. A new data point is added to the root of the tree by merging with the sub-cluster that has the smallest radius after merging. If the sub-cluster has child nodes, this merging process continues recursively down the tree. However, if the radius of the sub-cluster exceeds a threshold, and the number of sub-clusters is greater than a branching factor, the node divides. We find that radius threshold $=0.2$ yields the highest classification accuracy. We adopt branching factor $=50$ following the default values suggested by \textsc{scikit-learn}.
    \item \textbf{Spectral Clustering} starts by computing the Laplacian matrix ${\bf L}\equiv {\bf D}-{\bf A}$, where $\bf A$ is the affinity matrix and $\bf D$ is a diagonal matrix, \(D_{ii}\equiv\sum_j A_{ij}\). The affinity matrix effectively describes the ``similarity'' between data points. As suggested by \textsc{scikit-learn}, we calculate the affinity matrix using the radial basis function (RBF) kernel. Then, we determine the eigenvectors of $\bf L$ and construct a new matrix comprising the smallest eigenvectors. The number of eigenvectors is a free parameter. We find that the optimal number is 2, which equals the number of groups in our case. Each row in the new matrix defines the feature of a graph node. To cluster these features, we then employ an existing clustering method. For this purpose, we use the K-means method, a widely used approach for Spectral Clustering and the default option in \textsc{scikit-learn}.
\end{itemize}

\begin{table}
    \caption{Configurations of variables producing the highest accuracy for each classification method and number of dimensions (input variables). We list only the top two sequences from Agglomerative Clustering and BIRCH methods. We obtain these sequences by sequentially remove/add one variable each time based on the best configuration (highlighted in boldface) of each method, imposing lower thresholds for minimum accuracy. We do not show the cases with more than six variables as the accuracy decreases with more inputs.}
    \label{tab:configurations}
    \centering
    \renewcommand\arraystretch{1.4} 
    \begin{tabular}{cp{3.7cm}ccccc}
    \hline\hline
    Dim. & Best configuration & Method & Median acc.  \\
    \hline
    2
    & $\rapo,\Jr$ & Agglom. & 81.4\% \\
    \hline
    3
    & $\rapo,\Jr,\feh$ & Agglom. & 85.3\% \\
    & $r,\E,\feh$ & BIRCH & 81.5\% \\
    \hline
    4
    & $\rapo,\Jr,\feh,r$ & Agglom. & 82.9\% \\
    & $r,\E,\feh,\varepsilon$ & BIRCH & 81.5\% \\
    \hline
    5
    & $\rapo,\Jr,\feh,r,\varepsilon$ & Agglom. & 85.0\% \\
    & $r,\E,\feh,\varepsilon,\Jp$ & BIRCH & \textbf{89.4\%} \\
    \hline
    6
    & $\rapo,\Jr,\feh,r,\varepsilon,\age$ & Agglom. & \textbf{88.9\%} \\
    & $r,\E,\feh,\varepsilon,\Jp,\Jr$ & BIRCH & 86.4\% \\
    \hline
    \hline
    \end{tabular}
    \vspace{2mm}
\end{table}

\begin{figure*}
    \centering
    \includegraphics[width=\linewidth]{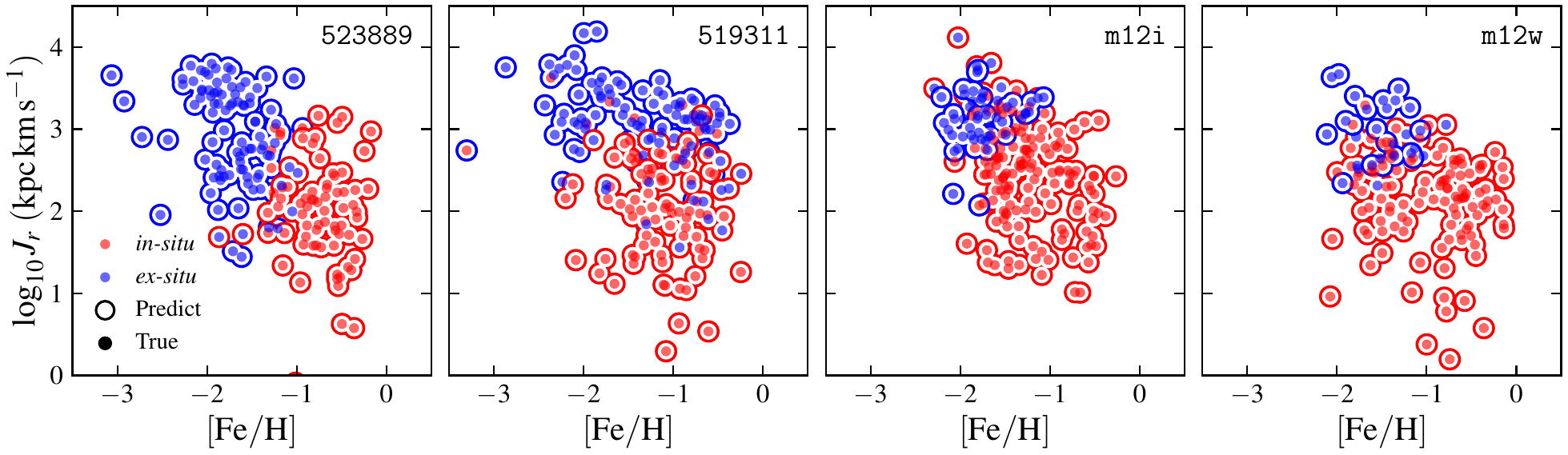}
    \vspace{-3mm}
    \caption{Classification results of \textit{in-situ} vs. \textit{ex-situ} presented in the $\feh$--$\Jr$ space for the four sample galaxies. The configuration is $\feh,\Jp,\varepsilon,r,$ and $\E$ using BIRCH. The colors of filled dots represent the true labels, while those of the open circles stand for the predicted progenitors.}
    \label{fig:feh_logJr}
\end{figure*}

Many of the clustering methods rely on the definition of the distance between data points. In our analysis, we adopt the Euclidean distance ($L^2$-distance) following the default setting in \textsc{scikit-learn}. Our tests indicate that other distance metrics, ranging from Manhattan distance ($L^1$) to Chebyshev distance ($L^\infty$), generally yield worse results compared to the Euclidean distance. Additionally, the concept of distance is influenced by the unit/scale of each variable. Given that different variables have various units that are not directly comparable, we standardize the data to ensure a zero mean and unit variance: $x\rightarrow x'\equiv(x-\bar{x})/\sigma$. This linear transformation equalizes the range across all variables.

Furthermore, some of these methods have many free parameters. However, the default parameters suggested by \textsc{scikit-learn} usually yield results that are comparable to or better than other parameter choices. Thus, we adhere to these default values unless specifically mentioned.

Although these clustering methods are in principle able to work in multi-dimensional spaces, they often encounter difficulties when applied to high-dimensional data. This is primarily due to the sparsity of data and the equalization of distances \citep{van_den_bussche_surprising_2001}. This challenge is commonly referred to as the \curse\ \citep[see a review by][]{domingos_few_2012}, referring to various problems that arise when analyzing data in high-dimensional spaces, which often do not occur in lower dimensions.

To understand and overcome the \curse, we conduct an analysis using subsets of the 10 variables. To identify which subsets of the 10 produce the most reliable and insightful classification between \textit{in-situ} and \textit{ex-situ} clusters, we carry out a comprehensive examination of all possible combinations of these variables. Given the 10 GC properties to consider, there are $(2^{10}-1)=1023$ unique configurations in total. 

We set the number of groups $N_{\rm g}$ to 2, corresponding to the \textit{in-situ} and \textit{ex-situ} populations of GCs. The clustering algorithms only divide the data into two groups, without specifying which one is \textit{in-situ} or \textit{ex-situ}. To correctly categorize these groups, we start by randomly assigning one group as \textit{in-situ} and the other as \textit{ex-situ}, and then compute the accuracy of such an assignment. The classification accuracy is defined as the proportion of clusters correctly assigned to their actual groups. Subsequently, we switch the labels of the two groups and calculate the accuracy again. The assignment with the higher accuracy is selected as the final classification. Using this methodology, the classification accuracy is guaranteed to be at least 50\%. In Appendix~\ref{sec:accuracy}, we rigorously prove this inequality in a more general case.

For each classification method, we scan the 1023 configurations to obtain the best one with the highest median accuracy over the four model galaxies. We define the median value as the mean of the median two galaxies if there are even number of galaxies. For most methods, the best configuration has $3-6$ variables. Including more variables does not always improve the performance, as stated by the \curse. Starting from the best configuration, we sequentially remove/add one variable each time to get the configurations yielding the highest accuracy with fewer/more number of input variables. This process produces a consistent sequence of variables with ranked significance. In the \textit{left panel} of Fig.~\ref{fig:dim_acc_no_threshold}, we show the median accuracy among the four sample galaxies for this sequence as a function of the number of variables. The highest median accuracy of Agglomerative Clustering and BIRCH is around 89\%. As demonstrated later in Appendix~\ref{sec:max_accuracy}, such a performance is close to the theoretical upper bound of a statistically significant accuracy $\approx90\%$.

Apart from producing high median accuracy, a robust configuration should avoid the unwanted scenario where the method performs well for some galaxies but yields substantially poorer accuracy for the remaining ones. To assess the robustness of the best configurations, we plot in the \textit{right panel} of Fig.~\ref{fig:dim_acc_no_threshold} the maximum and minimum accuracy over the sample galaxies for the same sequence of variables. Some configurations yield significantly low minimum accuracy close to 50\%, corresponding to a failed classification. To avoid these cases, we additionally impose a lower threshold for the minimum accuracy. We then repeat the same selection of best configurations but requiring the accuracy for any of the sample galaxies to exceed 70\%. Still, we find that the best configurations for Agglomerative Clustering and BIRCH produce median accuracy $\approx 89\%$, as presented in Tab.~\ref{tab:configurations}. Particularly, the BIRCH method achieves the highest median accuracy at 89.4\% with five variables: $r,\E,\feh,\varepsilon,\Jp$. In Fig.~\ref{fig:feh_logJr}, we illustrate the classification in the $\feh$--$\Jr$ space for the four sample galaxies using this configuration. Among them, the highest accuracy exceeds 90\% for \texttt{523889}, while the lowest accuracy still exceeds 80\% for \texttt{519311}. However, most other configurations fail to achieve a higher accuracy for this galaxy, suggesting a notable variation in the efficiency of identifying \textit{in-situ} vs. \textit{ex-situ} GCs across different galaxies. In Appendix~\ref{sec:max_accuracy}, we quantitatively demonstrate such a galaxy-to-galaxy variation by calculating the maximum achievable accuracy for the four galaxies. We notice that the \textit{in-situ} clusters misclassified as \textit{ex-situ} by this classification are more metal-rich than the general \textit{ex-situ} population. Additionally, these clusters are dynamically hotter than the remaining \textit{in-situ} GCs, characterized by higher $J_r$. These characteristics closely correspond to the ``Splash'' stellar population identified by \citet{belokurov_biggest_2020}. The ``Splash'' is believed to have originated in the MW's protodisk and was likely heated by the GS/E merger, resulting in kinematics that resemble those of the \textit{ex-situ} group. The impact of such early significant mergers will be further studied in \S\ref{sec:evolution}.

\begin{figure*}
    \centering
    \includegraphics[width=\linewidth]{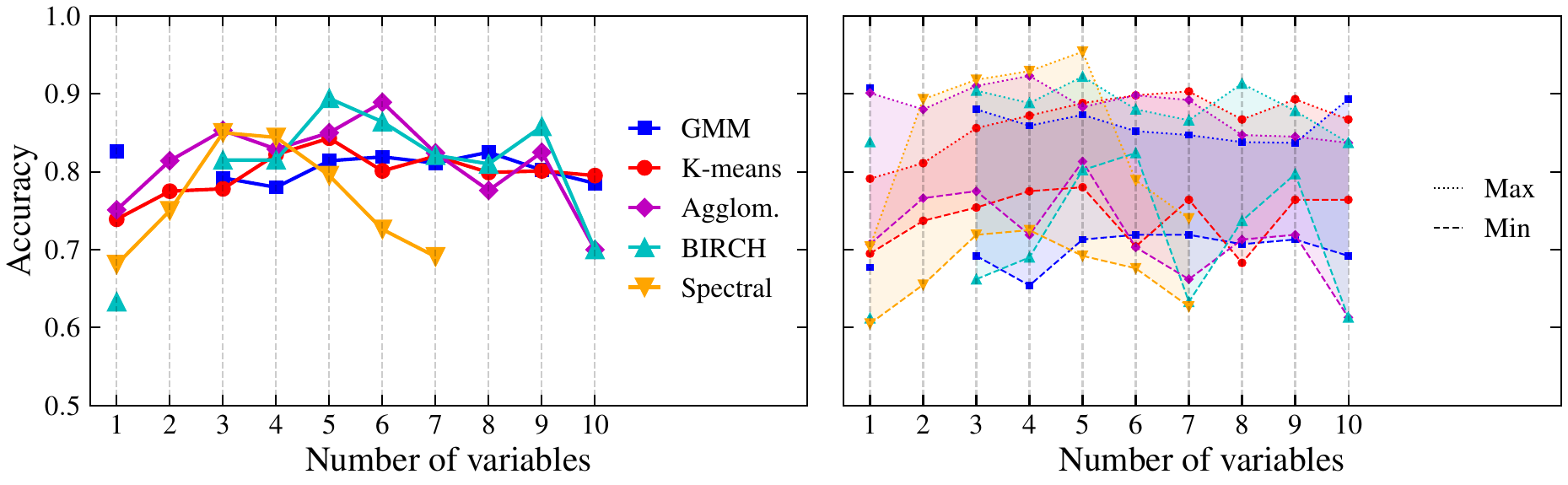}
    \vspace{-4mm}
    \caption{As Fig.~\ref{fig:dim_acc_no_threshold}, but we enforce the minimum accuracy of the four galaxies to exceed 70\% when selecting the optimal variable configuration. For the sub-optimal configurations with fewer/more inputs, we release the threshold to 60\% to allow more qualified configurations. If no configuration meets such a criterion, we skip this number of variables.}
    \label{fig:dim_acc}
\end{figure*}

Next, we sequentially remove/add variables to obtain a sequence of best configurations with fewer/more number of input variables. In this step, we release the lower threshold to 60\% to allow more qualified configurations. We present this sequence in Tab.~\ref{tab:configurations} up to six variables for Agglomerative Clustering and BIRCH. It is unnecessary to go beyond this number as the performance decreases with more variables. In Fig.~\ref{fig:dim_acc}, we plot the median, maximum, and minimum accuracy over the four sample galaxies for the sequence of variables selected for each method.

We notice that $\feh$ is an important property for categorizing \textit{in-situ} and \textit{ex-situ} GCs as it has high rankings in both sequences for the two best methods, Agglomerative Clustering and BIRCH. $\Jr$ and $r$ are other significant variable. However, since the two variables have strong correlation, having them together does not significantly increase the amount of information and has limited improvement in performance. Therefore, both methods only put high ranking to one of them but lower ranking to another. The circularity parameter $\varepsilon$ is also important for both methods but with lower rankings. In contrast, $\Jz$ and $\rperi$ appear to be not important for identifying \textit{in-situ} from \textit{ex-situ} GCs. 

\begin{table}
    \caption{Configurations of properties producing the highest accuracy using the hybrid approach combining dimensionality reduction and clustering methods. The first and second blocks show the PCA and t-SNE dimensionality reduction methods, respectively. For each method, we present the top two best configurations and clustering methods with the highest median accuracy among the four sample galaxies, with minimum accuracy exceeding 70\%.}
    \label{tab:configurations_pca}
    \centering
    \renewcommand\arraystretch{1.4} 
    \begin{tabular}{clccccc}
    \hline\hline
    Dim. reduction & Best configuration & Method & Median acc.  \\
    \hline
    PCA ($5\rightarrow3$)
    & $\age,\rperi,\rapo,\Jr,\varepsilon$ & Agglom. & \textbf{89.4\%} \\
    & $\age,\E,\Jp,\Jr,\feh$ & BIRCH & 88.1\% \\
    \hline
    t-SNE ($5\rightarrow2$)
    & $\E,\rapo,\Jr,\feh,\varepsilon$ & GMM & \textbf{87.5\%} \\
    & $\E,\Jr,\feh,\rapo,\varepsilon$ & BIRCH & 86.9\% \\
    \hline
    \hline
    \end{tabular}
    \vspace{2mm}
\end{table}

\subsection{Dimensionality reduction}
\label{sec:dim_reduce}

Since the \curse\ greatly limits the performance of classification methods in high-dimensional spaces, we also implement a two-step process that incorporates a dimensionality reduction step before classification to overcome such a limitation and to learn valuable information from more input variables. Dimensionality reduction projects the original data into lower-dimensional sub-spaces. The projected coordinates are either linear or non-linear combinations of the original variables. The primary goal of dimensionality reduction is to identify the most significant combinations that preserve the most information of the original data.

We employ here two dimensionality reduction methods.
\begin{itemize}[labelindent=0pt,labelwidth=10pt,labelsep*=0pt,leftmargin=!,align=parleft]
    \item \textbf{Principal Component Analysis (PCA)} is a linear method for dimensionality reduction. It aims to maximize variance within lower-dimensional sub-space. To achieve this aim, PCA identifies and selects the first few eigenvectors of the covariance matrix for the original data. These eigenvectors are then used as the coordinates for the lower-dimensional sub-space. This approach effectively captures the most significant linear combinations in the original high-dimensional data.
    \item \textbf{t-Distributed Stochastic Neighbor Embedding (t-SNE)} is a non-linear dimensionality reduction technique developed by \citet{van_der_maaten_visualizing_2008}. This method begins by computing the affinity matrix in the original high-dimensional space using a Gaussian kernel. The bandwidth of this kernel is adaptive based on the density of data points, such that the Shannon entropy equals the $\log_2$ of a threshold parameter called perplexity. By construction, perplexity is constrained not to exceed the number of data points and is commonly set higher for more data. In our analysis, perplexity $=50$ is found the most effective. On the other hand, the affinity in the lower-dimensional sub-space is calculated using a Student t-distribution with one degree of freedom. The method then maps the high-dimensional data to lower-dimensional sub-space while minimizing the \cite{kullback_information_1951} divergence between the two sets of affinities. For our work, we use the t-SNE implementation by the \textsc{scikit-learn} package.
\end{itemize}

To assess this two-step procedure, we apply the two dimensionality reduction methods to all variable configurations, subsequently reducing them to an $M$-D sub-space. For the purpose of identifying \textit{in-situ} and \textit{ex-situ} GC populations, PCA is found to be the most effective at $M=3$, yielding a highest median accuracy of $89.4\%$. On the other hand, t-SNE shows optimal performance at $M=2$, with a median accuracy of $87.5\%$. The top two best configurations for these two methods are listed in Tab.~\ref{tab:configurations_pca}. Compared to the direct clustering approach, both PCA and t-SNE do not improve the highest median accuracy of the best clustering methods. 

\begin{figure*}
    \centering
    \includegraphics[width=\linewidth]{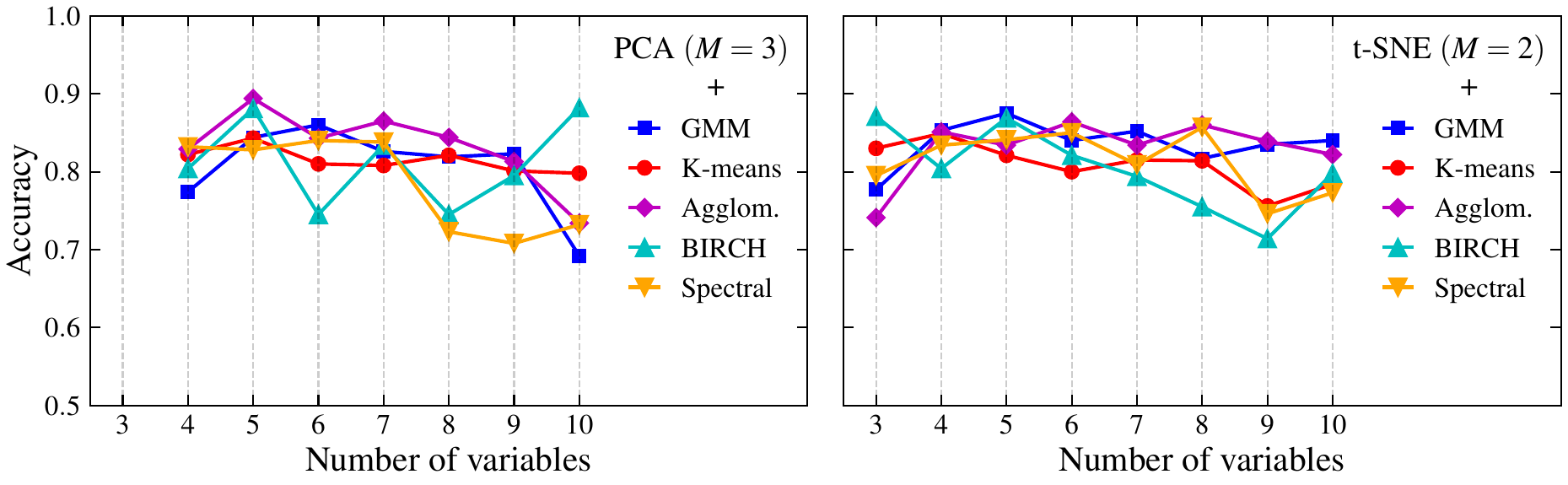}
    \vspace{-4mm}
    \caption{As the \textit{left panel} of Fig.~\ref{fig:dim_acc}, but with PCA ($M=3$, \textit{left}) and t-SNE ($M=2$, \textit{right}) dimensionality reduction before applying the clustering methods.}
    \label{fig:dim_acc_reduce}
\end{figure*}

Starting from the best configurations, we following the same process in \S\ref{sec:direct_clustering} to get the consistent sequence of variables for each method. We plot the median accuracy of such a sequence over the four sample galaxies in Fig.~\ref{fig:dim_acc_reduce}. Compared with Fig.~\ref{fig:dim_acc}, while dimensionality reduction methods do not enhance the performance of the best clustering methods, they significantly boost the effectiveness of other less optimal clustering approaches like GMM and Spectral Clustering, narrowing the performance gap between various methods. This is likely because the dimensionality reduction methods regulate the input data into a sub-space where clustering features are more pronounced. This enables different clustering methods to produce more coherent results, as well as enhance their overall performance. 

\subsection{Supervised classification}
\label{sec:supervised_classification}

To address the \curse, another effective strategy is to employ supervised classification methods. These methods train a classifier in multi-dimensional space using a set of training data. To obtain a reliable training set, we first apply a probabilistic clustering method to estimate the probability of each data point belonging to either the \textit{in-situ} or \textit{ex-situ} category. This step is conducted in lower-dimensional spaces to improve efficiency. We then select unambiguous data points to form the training set, where an unambiguous data point has a probability greater than 90\% of being either \textit{in-situ} or \textit{ex-situ}. The supervised classifier is trained to recognize the distribution of these clearly-defined training data within the full 10D space. We then predict the classifications for all data points.

We have explored various thresholds for determining unambiguous data. A lower threshold of 50\% fails to accurately label training data near the boundary between groups, while a higher threshold of 99\% leads to insufficient number of data points in the training set, failing to capture the defining characteristics of the two populations. Both scenarios reduce the classification accuracy compared to our selected threshold of 90\%. 

The two-step classification procedure requires the unsupervised method to be predictive in estimating the probability of data points. Among all the unsupervised methods we have considered, only GMM has the ability of probabilistic prediction. Consequently, we choose GMM as the first step of the procedure. For the second step of supervised classification, we consider the following popular and well-established methods. These supervised classification algorithms are also implemented in the \textsc{scikit-learn} package. 

\begin{itemize}[labelindent=0pt,labelwidth=10pt,labelsep*=0pt,leftmargin=!,align=parleft]
    \item \textbf{Support Vector Classification (SVC)} focuses on identifying hyperplanes in multi-dimensional space to divide data into distinct regions. The hyperplane maximizes the separation between the two regions it divides. This separation is defined as the distance between the hyperplane and the nearest training data point (i.e., support vector) on either side of the hyperplane. SVC ensures that the hyperplane not only separates the regions effectively but also provides the greatest possible distinction between them.
    \item \textbf{K-Nearest Neighbors (K-NN) Classification} builds the classifier based on a given training set. The core principle of K-NN is to classify new data points by identifying which group has the majority among its K closest neighbors in the training set. The parameter K, representing the number of neighbors to consider, is adjustable and plays a crucial role in the classification process. A smaller K often leads to overfitting, while a larger K tends to underfit the data. In our study, we examine various values of K from 1 to 10 to determine the optimal choice. In most case, K $=5$ yields the best classification accuracy.
    \item \textbf{Gaussian Process Classification (GPC)} is a probabilistic approach that uses Gaussian process models for classification. It places a Gaussian process prior to a latent function, which is then converted to a classification probability $\in[0,1]$ using a logistic function. GPC classifies new data points based on their similarity to the training set via a kernel function. In this work, we use the RBF kernel, following default setting in \textsc{scikit-learn}. The hyper-parameters of the kernel are optimized during fitting. We also evaluate alternative kernels such as the Mat\'{e}rn and the rational quadratic kernels. They do not demonstrate significantly different results compared to the RBF kernel.
\end{itemize}

However, the supervised classification methods do not surpass the accuracy achieved by direct clustering methods. The highest accuracy is 81.9\% obtained by SVC, with input variables of $\feh,\Jr,\Jp,\Jz,r,$ and $\E$. This accuracy is 8\% lower than the best performance achieved using direct clustering methods (Tab.~\ref{tab:configurations}) and the dimensionality reduction approach (Tab.~\ref{tab:configurations_pca}).

Moreover, if we place an additional dimensionality reduction step prior to GMM, similarly to \S\ref{sec:dim_reduce}, the results still do not improve. The under-performance of this approach is likely because 1) the training set derived from GMM is too small to adequately sample the entire property space; 2) the probabilistic labels by GMM are not entirely accurate, leading to potential errors in the subsequent classification step.

\section{Identify progenitor galaxies}
\label{sec:progenitors}

In this section, we focus on the \textit{ex-situ} component, which is comprised of groups of GCs accreted from past mergers. Our goal is to evaluate classification methods on model galaxies to distinguish \textit{ex-situ} GCs from distinct progenitor galaxies, which is the key topic in galactic archaeology. 

We consider only progenitor galaxies contributing a minimum of three surviving GCs. This threshold is crucial as the multi-dimensional distribution becomes under-sampled for progenitors contributing less GCs. Under this criterion, we identify four, four (plus one \textit{early-accreted}), two (plus one \textit{early-accreted}), and four major progenitors for the model galaxies \texttt{523889}, \texttt{519311}, \texttt{m12i}, and \texttt{m12w}, respectively. These identified mergers represent key characteristics in the galaxy assembly histories and are significant for shaping the present-day properties of GC systems.

The ideal classification procedure should predict the correct number of mergers and accurately label GCs to their progenitors. However, this task is challenging particularly due to the various numbers of GCs each merger contributed. In the case of the MW and its model analogs, major mergers, such as the GS/E, are known to contribute $\sim 30$ GCs \citep{massari_origin_2019}. In contrast, smaller mergers contribute only a few. The GCs from the major mergers act as the noisy background when identifying smaller mergers, contaminating the GC distribution of the latter in the multi-dimensional property space. This limits clustering methods that rely on the density of data points or linkage distances. Such methods, which are effective in identifying dominant mergers, may not be as effective in distinguishing smaller ones.

To study the number of mergers that can be accurately identified, we explore a range of clustering groups from $N_{\rm g}=2$ to 5 in the following subsections. We interpret the $N_{\rm g}$ groups as the top $N_{\rm g}-1$ most significant mergers, with an additional group representing the remaining \textit{ex-situ} clusters. We also employ quantitative methods such as the information criterion and cross-validation techniques to objectively determine the number of groups that best represent the \textit{ex-situ} GC population.

\subsection{Two groups}
\label{sec:n_g_2}

First, we set the number of groups $N_{\rm g}$ to 2. Under this configuration, one group represents the most dominant merger event, which is the GS/E-analog (\texttt{b}). The other group stands for all remaining \textit{ex-situ} GCs. Algorithmically, this task is similar to \S\ref{sec:in_vs_ex}, but applied to \textit{ex-situ} clusters only.

To evaluate the classification results, we calculate the proportion of GCs in the two groups that are correctly labeled. Similarly to \S\ref{sec:in_vs_ex}, we initially calculate the accuracy of assigning one group to the most dominant merger and the other to the remaining \textit{ex-situ} GCs. Next, we switch the roles of the two groups and re-assess the accuracy. The assignment with more correctly labeled GCs is considered the correct one. By this definition, the accuracy is always at least 50\%. 

In practice, errors in categorizing \textit{in-situ} and \textit{ex-situ} populations, as assessed in \S\ref{sec:in_vs_ex}, can influence the accuracy of subsequent steps of identifying specific progenitors. However, to independently evaluate the effectiveness of progenitor identification for \textit{ex-situ} GCs, we proceed with an assumption that the exact labels of \textit{in-situ} GCs are known. By isolating this step from the initial categorization of \textit{in-situ} vs. \textit{ex-situ}, we ensure that our analysis is robust to prior biases and is tailored to optimize the efficiency for this particular purpose.

Next, we apply the aforementioned classification methods to the \textit{ex-situ} GCs. For the direct clustering approach, we list the best property configurations for each dimension in Tab.~\ref{tab:configurations_exsitu_2_groups}. We also employ the same 70\% threshold for the minimum accuracy to ensure the robustness of categorization. The highest accuracy is 90.6\% using Agglomerative Clustering or BIRCH. Different from the task of identifying \textit{in-situ} vs. \textit{ex-situ} clusters, we need one or two more variables to distinguish the more subtle differences between GCs originating from various progenitors.

\begin{figure*}
    \centering
    \includegraphics[width=\linewidth]{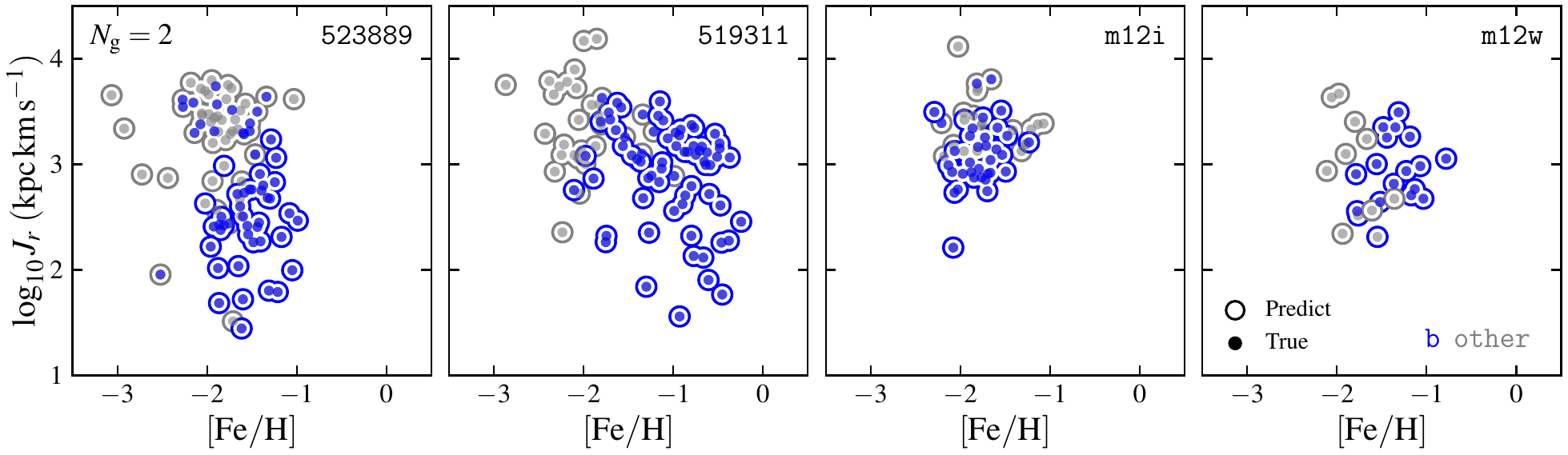} \\
    \vspace{2mm}
    \includegraphics[width=\linewidth]{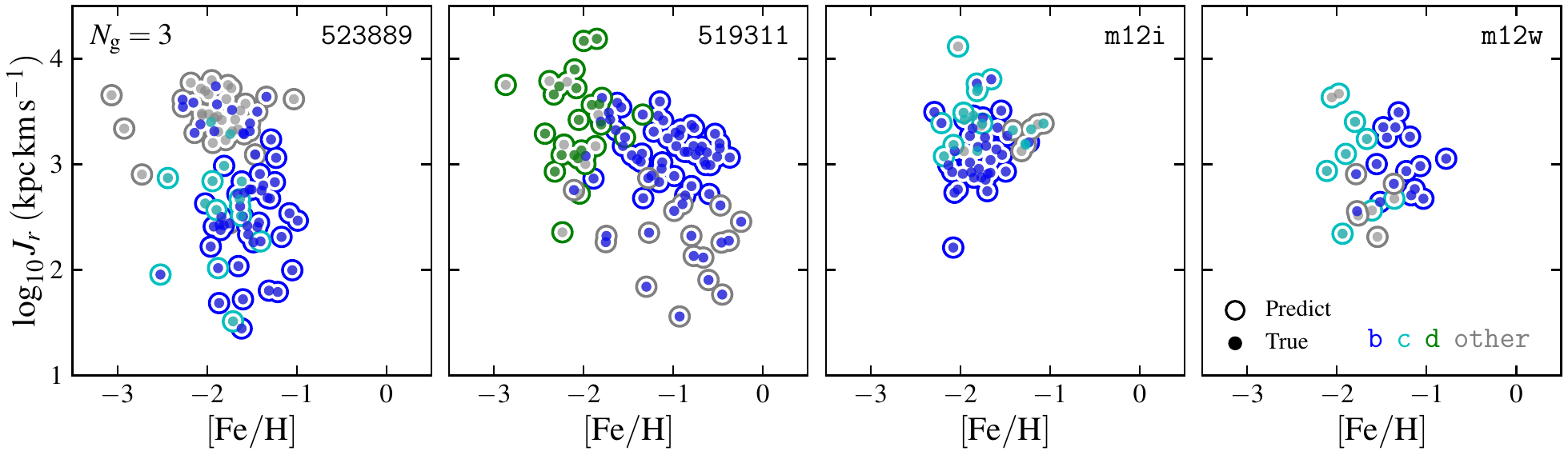} \\
    \vspace{2mm}
    \includegraphics[width=\linewidth]{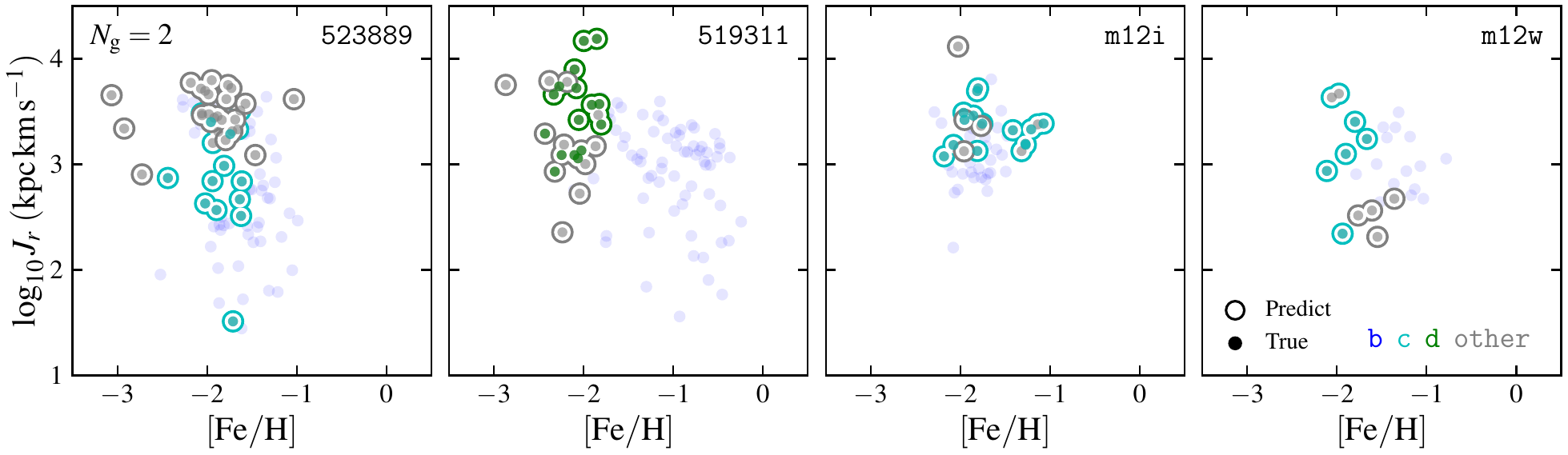} \\
    \vspace{0mm}
    \caption{Classification results for progenitors of \textit{ex-situ} GCs, with $N_{\rm g}=2$ (\textit{top row}) and 3 (\textit{middle row}) for all \textit{ex-situ} clusters, and $N_{\rm g}=2$ for \textit{ex-situ} clusters excluding the most dominant merger  (\textit{bottom row}). Although we visualize the results in the $\feh$--$\Jr$ space, the classification is performed with more variables. We use $\age,\Jz,\feh,\Jp,\rapo,\varepsilon,$ and $\E$ with Agglomerative Clustering for the \textit{top row}, $\feh,\varepsilon,\rapo,\E,$ and $\age$ with Agglomerative Clustering for the \textit{middle row}, and $\feh,\Jr,\Jz,\rperi,$ and $\E$ with K-means for the \textit{bottom row}. The colors of filled dots represent the true labels, while those of the open circles stand for the predicted progenitors. For reference, we also show the clusters from the most dominant merger as faint blue dots in the \textit{bottom row}.}
    \label{fig:more_groups}
\end{figure*}

In the \textit{top row} of Fig.~\ref{fig:more_groups}, we display the classification results for $N_{\rm g}=2$ in the $\feh$--$\Jr$ space. The galaxy-to-galaxy variation in distributions of the GS/E-analog vs. other \textit{ex-situ} GCs is pronounced due to the distinct assembly histories of different galaxies. For instance, in the case of \texttt{m12i}, the GS/E-analog and other \textit{ex-situ} GCs largely overlap in the $\feh$--$\Jr$ space alone, while the two groups are more separable in other model galaxies. Such a variance again emphasizes the need of more variables to accurately identify progenitors of GCs for all model galaxies. 

\begin{table}
    \caption{As Tab.~\ref{tab:configurations}, but for the purpose of identifying GS/E and other \textit{ex-situ} clusters.}
    \label{tab:configurations_exsitu_2_groups}
    \centering
    \renewcommand\arraystretch{1.4} 
    \begin{tabular}{cp{3.7cm}ccccc}
    \hline\hline
    Dim. & Best configuration & Method & Median acc.  \\
    \hline
    2
    & $\age,\Jz$ & Agglom. & 84.1\% \\
    & $\feh,\age$ & BIRCH & 75.4\% \\
    \hline
    3
    & $\age,\Jz,\feh$ & Agglom. & 81.3\% \\
    & $\feh,\age,\Jp$ & BIRCH & 83.1\% \\
    \hline
    4
    & $\age,\Jz,\feh,\Jp$ & Agglom. & 83.1\% \\
    & $\feh,\age,\Jp,\Jz$ & BIRCH & 83.1\% \\
    \hline
    5
    & $\age,\Jz,\feh,\Jp,\rapo$ & Agglom. & 83.1\% \\
    & $\feh,\age,\Jp,\Jz,\rapo$ & BIRCH & 83.1\% \\
    \hline
    6
    & $\age,\Jz,\feh,\Jp,\rapo,\varepsilon$ & Agglom. & 82.8\% \\
    & $\feh,\age,\Jp,\Jz,\rapo,\varepsilon$ & BIRCH & 82.8\% \\
    \hline
    7
    & $\age,\Jz,\feh,\Jp,\rapo,\varepsilon,\E$ & Agglom. & \textbf{90.6\%} \\
    & $\feh,\age,\Jp,\Jz,\rapo,\varepsilon,\E$ & BIRCH & \textbf{90.6\%} \\
    \hline
    \hline
    \end{tabular}
    \vspace{2mm}
\end{table}

We also explore both the dimensionality reduction and supervised classification approaches introduced in \S\ref{sec:dim_reduce} and \S\ref{sec:supervised_classification}. However, these methods do not enhance the performance compared to the direct clustering methods. Therefore, we do not delve deeper into these approaches.

\subsection{Three and more groups}
\label{sec:n_g_greater_than_3}

Now, we expand our analysis to include $N_{\rm g}=3$ groups, corresponding to the most dominant merger, the second most massive merger (\texttt{c} for \texttt{523889}, \texttt{m12i}, and \texttt{m12w}; and \texttt{d} for \texttt{513911}, where \texttt{513911-c} is \textit{early-accreted}), and all other \textit{ex-situ} clusters. In this three-group scenario, the lowest possible accuracy is $1/3$, indicating a completely random classification. We provide a rigorous proof of this inequality in Appendix~\ref{sec:accuracy}. 

For $N_{\rm g}=3$, we find no configuration that yields a minimum accuracy exceeding 70\%, regardless of the clustering methods used. After we relax this constraint to 65\%, several configurations emerge. The most effective configuration is $\feh,\varepsilon,\rapo,\E,$ and $\age$ with Agglomerative Clustering. The accuracy is 72.0\% for \texttt{523889}, 67.0\% for \texttt{519311}, 73.6\% for \texttt{m12i}, and 74.1\% for \texttt{m12w}. Although this approach successfully identify the most dominant merger, as in the $N_{\rm g}=2$ case, it tends to mis-classify the GCs from the second largest merger, as shown in the \textit{middle row} of Fig.~\ref{fig:more_groups}. It assigns either more GCs (\texttt{523889}) or less GCs (\texttt{m12i}) to the second largest merger. Also, the third group is poorly constructed for galaxies like \texttt{519311} and \texttt{m12w}. 

The mis-classification is likely due to the dominance of the GS/E-analog over others. In each of the sample galaxies, the most significant merger contributes more than half of the \textit{ex-situ} GC population. Such large mergers tend to create a dense and noisy region in the multi-dimensional property space, concealing the useful clustering features to identify the remainder groups.

In response to such a challenge, we explore an alternative approach to classify the \textit{ex-situ} population into three groups. This approach involves two recursive steps. The first step is identical to \S\ref{sec:n_g_2}, where we divide the \textit{ex-situ} population into two groups, representing the GS/E and the rest of the \textit{ex-situ} GCs. Secondly, we remove the GCs assigned to the GS/E from the input data. We then further divide the remaining \textit{ex-situ} GCs into two new groups, standing for the second largest merger and the other GCs. Since we have evaluated the first step in the previous subsection, here we proceed with the assumption that the exact labels for the GS/E GCs are already known. This allows us to study the performance of the second step independently from prior classification. That is to say, we assume a 100\% accuracy for the first step and thus overestimate the joint accuracy in the following analysis.

For this approach, we find an optimal configuration using $\feh,\Jr,\Jz,\rperi,$ and $\E$ with K-means clustering. This configuration achieves an accuracy of 81.6\% for \texttt{523889}, 72.0\% for \texttt{519311}, 84.2\% for \texttt{m12i}, and 81.8\% for \texttt{m12w}. The classification results are displayed in the \textit{lower row} of Fig.~\ref{fig:more_groups}.

Compared to the direct approach of splitting into three groups, this recursive method more accurately identifies GCs from the second largest merger. However, it is important to note that this does not necessarily indicate that this method outperforms the direct method, because of the potential contamination by GCs from the most dominant merger. Indeed, if we use the best configuration from \S\ref{sec:n_g_2} as the first step to exclude the most massive merger, the median accuracy falls below 70\%.

Again, the dimensionality reduction and supervised classification approaches do not improve the accuracy of direct clustering methods. We do not discuss them in this subsection.

Additionally, we do not present the results with $N_{\rm g}=4$ and higher because neither the direct approach nor the recursive approach yields valid classifications in these cases. 
These secondary mergers typically contribute far fewer GCs than the most dominant ones, being insufficient to form a distinct core in the multi-dimensional property space that stands out from the background of dominant mergers. Any clustering method that relies on the density or linkage distance tends to fail in these cases.

Furthermore, the assembly histories of MW-like galaxies show that most mergers occurred at around the same time or earlier the GS/E (see Fig.~\ref{fig:mh_log_vs_tlookback}; these galaxies experienced no major mergers after the GS/E by construction). The GS/E-analog can significantly perturb the orbits of previously accreted \textit{ex-situ} GCs, making them less distinguishable in the property space. We provide an in-depth analysis on the impact of major mergers in \S\ref{sec:evolution}. 

We quantitatively study the optimal number of groups that represents the \textit{ex-situ} population using the Bayesian Information Criterion (BIC). The GMM is particularly suited for this analysis, as it employs a likelihood-maximization method to fit Gaussian components, where the BIC can be unambiguously defined as:
\begin{equation*}
    {\rm BIC} \equiv -2\ln{{\cal L}_{\rm max}} + k \ln{n}
\end{equation*}
where ${\cal L}_{\rm max}$ is the maximized likelihood given by Eq.~(\ref{eq:gmm}), $n$ is the number of data points, and $k$ is the number of model parameters. In the case of $N_{\rm g}$-component GMM, 
\begin{equation*}
    k = N_{\rm g} N (N + 1) / 2 + N_{\rm g} N + (N_{\rm g} - 1)
\end{equation*}
in which $N$ is the number of input variables. The first term counts the independent elements in the covariance matrix, the second term represents the number of mean values, and the last term is the number of independent weights.

The BIC is useful for model selection. A lower BIC value generally indicates better model, considering both the fit to data and the number of parameters. In general, greater number of groups increases the likelihood, therefore reducing the first terms of BIC. However, the second term puts penalty on the number of parameters, preventing $N_{\rm g}$ from growing out of control. 

\begin{figure}
    \centering
    \includegraphics[width=1\linewidth]{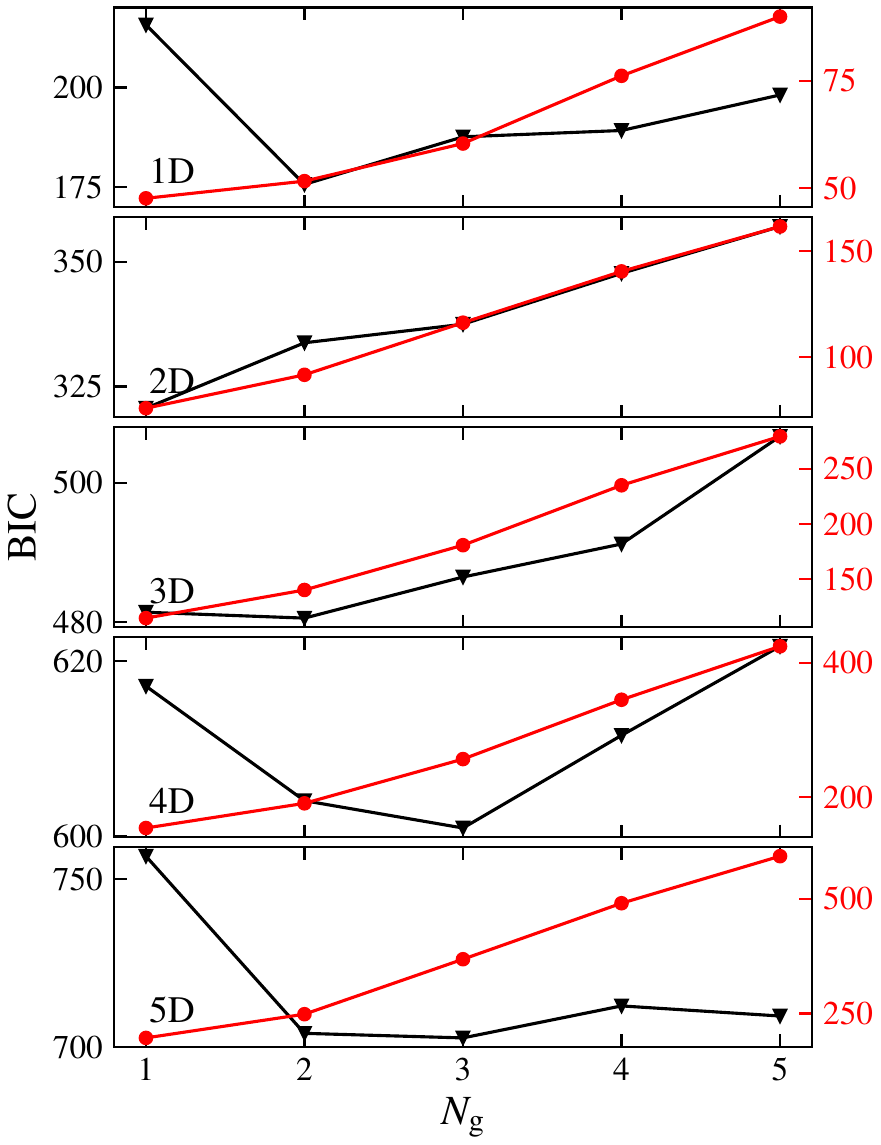}
    \vspace{-2mm}
    \caption{Bayesian information criterion for GMM as a function of number of groups $N_{\rm g}$. The black curve (\textit{left axis}) represents BIC on the full data, while the red curve (\textit{right axis}) is for the validation sets using five-fold cross-validation. For each row, we fix the number of input variables and compute the median BIC of all property configurations on the four galaxies. We show the two BIC versions separately on two axes because they span distinct ranges. A lower BIC indicates better fit to the data.}
    \label{fig:gmm_ic_vc}
\end{figure}

In Fig.~\ref{fig:gmm_ic_vc} we plot the median BIC for various configurations of variables with $N_{\rm g}=1-$ 5. We analyze the configurations from 1D to 5D separately. We do not plot the BIC for higher dimensions, where the \curse\ becomes important. We find that the BIC reaches its minimum at $N_{\rm g}=1-3$. Greater numbers of groups are not favored as they do not significantly improve the model's ability to fit the data while adding a large amount of parameters.

In addition to directly using the BIC, we also employ cross-validation to assess the optimal number of groups. The primary aim of cross-validation is to prevent overfitting, which can occur when a model has too many parameters. The cross-validation process splits the data into a training set and a validation set. The model parameters are fitted using the training set, while the model's performance is evaluated using the validation set. For the GMM specifically, we calculate the BIC for the validation set based on the model trained with the training set.

We adopt a five-fold cross-validation approach, where we divide the data into five sets. Each set takes turns to serve as the validation set, while the remaining four are combined as the training set. We present the mean values of the five rounds of calculations in Fig.~\ref{fig:gmm_ic_vc}, where the BIC of the training sets increases monotonically with $N_{\rm g}$. Such an increasing trend suggests that the problem of overfitting becomes more severe with a larger number of groups. In the case of identifying progenitors of \textit{ex-situ} GCs, the overfitting often leads to mis-classification. Therefore, using too many parameters not only leads to inefficient modeling (as suggested by BIC) but also causes problematic representation of the entire data set (as suggested by cross-validation).

\section{Application to the Milky Way globular clusters}
\label{sec:mw}

In this section, we employ the classification methods we have evaluated earlier to the real Galactic GCs. Before delving deep into the classification process, we first introduce the observational data we adopt for the MW GCs. The 3D positions and 3D velocities of the MW GCs are from the fourth edition of the \citet{hilker_galactic_2019} catalog\footnote{\url{https://people.smp.uq.edu.au/HolgerBaumgardt/globular/}}, which contains more than 160 Galactic GCs. Utilizing this catalog, we calculate $\Jr$, $\Jp$, $\Jz$, $\varepsilon$, $r$, $\rperi$, $\rapo$, and $\E$. These calculations are performed using the \textsc{agama} software \citep{vasiliev_agama_2019}, assuming an analytical MW potential from \cite{belokurov_energy_2023}. This potential model resembles the three-component model \texttt{MWPotential2014} by \citet{bovy_galpy_2015}. However, it increases the virial mass of the Navarro-Frenk-White halo component to $10^{12}\Msun$, in agreement with the latest measurement of the MW's halo mass derived from stellar streams \citep{ibata_charting_2023}. Additionally, the concentration parameter is adjusted to 19.5 to match the circular velocity at the Solar radius. However, different choices of the potential model only slightly alter the classification results as long as the model is consistent with the observational constraints.

The metallicities of the MW GCs are from the 2010 edition of the catalog by \citet{harris_catalog_1996}\footnote{\url{https://physics.mcmaster.ca/~harris/mwgc.dat}}. This catalog provides metallicity measurements for 152 Galactic GCs. We cross match this catalog with the \citet{hilker_galactic_2019} catalog, resulting in a set of 150 clusters. Two clusters in the \citet{harris_catalog_1996} catalog, BH 176 and GLIMPSE02, do not appear in the \citet{hilker_galactic_2019} catalog.

Unfortunately, a large number of MW GCs lack reliable age measurements. Due to this limitation, we decided not to include age as an input in our analysis here. 

Throughout this section, we compare our classification results with \citet{massari_origin_2019} and \citet{belokurov_-situ_2024}. \citet{massari_origin_2019} classified the progenitors of 151 Galactic GCs, in which 146 are included in our sample. They considered mainly the kinematics of GCs, including the apocenter radius, maximum height above the disk, orbital circularity, specific angular momentum along and perpendicular to the rotational axis, and specific energy. Instead of using unsupervised clustering techniques, their classification is based largely on prior knowledge of the progenitor galaxy properties. For instance, their criteria to identify \textit{in-situ} clusters are based on a selected GC sample on the in-situ branch of the age--metallicity relation. Their classification is presented as rectangular regions in these property spaces. If a GC is located in the overlapping region of two rectangles, they refer to this GC as a tentative candidate of both progenitors. We refer to a GC ``unambiguous'' if these authors assigned the GC to only one unique progenitor.

\citet{belokurov_-situ_2024} classified \textit{in-situ} vs. \textit{ex-situ} origins for 165 Galactic GCs with measured kinematics using the same \citet{hilker_galactic_2019} catalogue as in our work. They also applied the same gravitational potential to compute the energy of GCs. Therefore, we can directly compare our classification results with theirs in the IoM space, where these authors used an empirical boundary to separate the \textit{in-situ} and \textit{ex-situ} populations, see their Eq.~(1). This boundary was carefully selected to isolate \textit{in-situ} and accreted field stars within the range of metallicities where their classification is robust. Although the two GC populations have distinct distributions in other parameter spaces, such as the $\rm[Mg/Fe]$--$\rm[Al/Fe]$ and age--metallicity planes, the boundary in the $E-L_z$ space provides a more clear separation between the two populations. They showed that using machine learning methods cannot substantially improve the classification accuracy in the $E-L_z$ space.

We follow the same pipeline used in our previous analysis to classify the MW GCs. That is, first we categorize GCs into \textit{in-situ} and \textit{ex-situ} groups and then identify progenitors of \textit{ex-situ} GCs. Given that direct clustering methods (\S\ref{sec:direct_clustering}) generally perform comparably or better to the dimensionality reduction approach (\S\ref{sec:dim_reduce}) and the hybrid methods combining clustering and supervised classification (\S\ref{sec:supervised_classification}), we concentrate solely on direct clustering methods in this section. 

We use a novel Bayesian approach to combine results from various clustering methods. Each configuration of GC properties, associated with a specific clustering method, forms a unique ``classifier''. There are $5\times(2^{9}-1)=2555$ classifiers, excluding the age variable. To ensure the reliability of the combined results, we only consider classifiers with a minimum accuracy greater than 78\%. 
However, we have tested that setting a different threshold from 50\% to 80 \% only influences the assignment of 3 GCs. This threshold is selected just to ensure the number of valid classifiers $N_{\rm c}$ is around $10-100$. Too few classifiers leads to under-representation, while too many classifiers may bias the classification due to the large number of correlated classifiers, as discussed later.

We combine the classification results using Bayes' theorem,
\begin{equation}
    P_{\rm post}(z_i=j) = \frac{P_{\rm prior}(z_i=j) \prod_{k=1}^{N_{\rm c}} P(k|z_i=j)}{P_{\rm marginal}}
    \label{eq:Bayesian}
\end{equation}
where $z_i\in\{$\textit{in-situ}, \textit{ex-situ}$\}$ stands for the origin of the $i$-th data point, and $P_{\rm prior}$ is the prior probability which is assumed to be 0.5. The denominator
\begin{align*}
    P_{\rm marginal}\equiv &P_{\rm prior}(z_i=j)\prod_{k=1}^{N_{\rm c}}P(k|z_i=j) \\
    + &P_{\rm prior}(z_i\neq j)\prod_{k=1}^{N_{\rm c}}P(k|z_i\neq j)
\end{align*}
denotes the marginal probability. The products stand for the conditional probability of $i$-th data point being classified \textit{in-situ}/\textit{ex-situ} by all classifiers $k=1,2,\cdots,N_{\rm c}$, if it is indeed \textit{in-situ}/\textit{ex-situ}. We assume the independence of classifiers so that the joint probability equals the product of individual conditional probabilities, which is given by
\begin{equation*}
    P(k|z_i=j)= \left\{
    \begin{array}{ll}
        P_k, & \text{if}\ z_i^k=j  \\
        1-P_k, & \text{if}\ z_i^k\neq j 
    \end{array}
    \right. 
\end{equation*}
in which $z_i^k$ is the classification result of the $i$-th data point by the $k$-th classifier, and $P_k$ is the accuracy of the classifier. To label the groups as either \textit{in-situ} or \textit{ex-situ}, we define the group with greater mean metallicity and mean circularity, and lower mean energy as \textit{in-situ}. 

Note that when the accuracy of a classifier $P_k=0.5$, it can be factored out from both the numerator and denominator of Eq.~(\ref{eq:Bayesian}). Therefore, classifiers with an accuracy of 0.5 do not contribute to the calculation of posterior accuracy. This is consistent with our definition of accuracy, which is greater than or equal to 0.5 by construction (see Appendix~\ref{sec:accuracy} for the case of $N_{\rm g}=2$).

We use the posterior probability derived from the Bayesian framework to define \textit{in-situ} vs. \textit{ex-situ} origins. Specifically, clusters with the posterior probability $P_{\rm post}(z_i=\textit{in-situ})>0.5$ are classified as \textit{in-situ}. The remaining clusters are categorized as \textit{ex-situ}. 

\begin{figure*}
    \centering
    \includegraphics[width=0.8\linewidth]{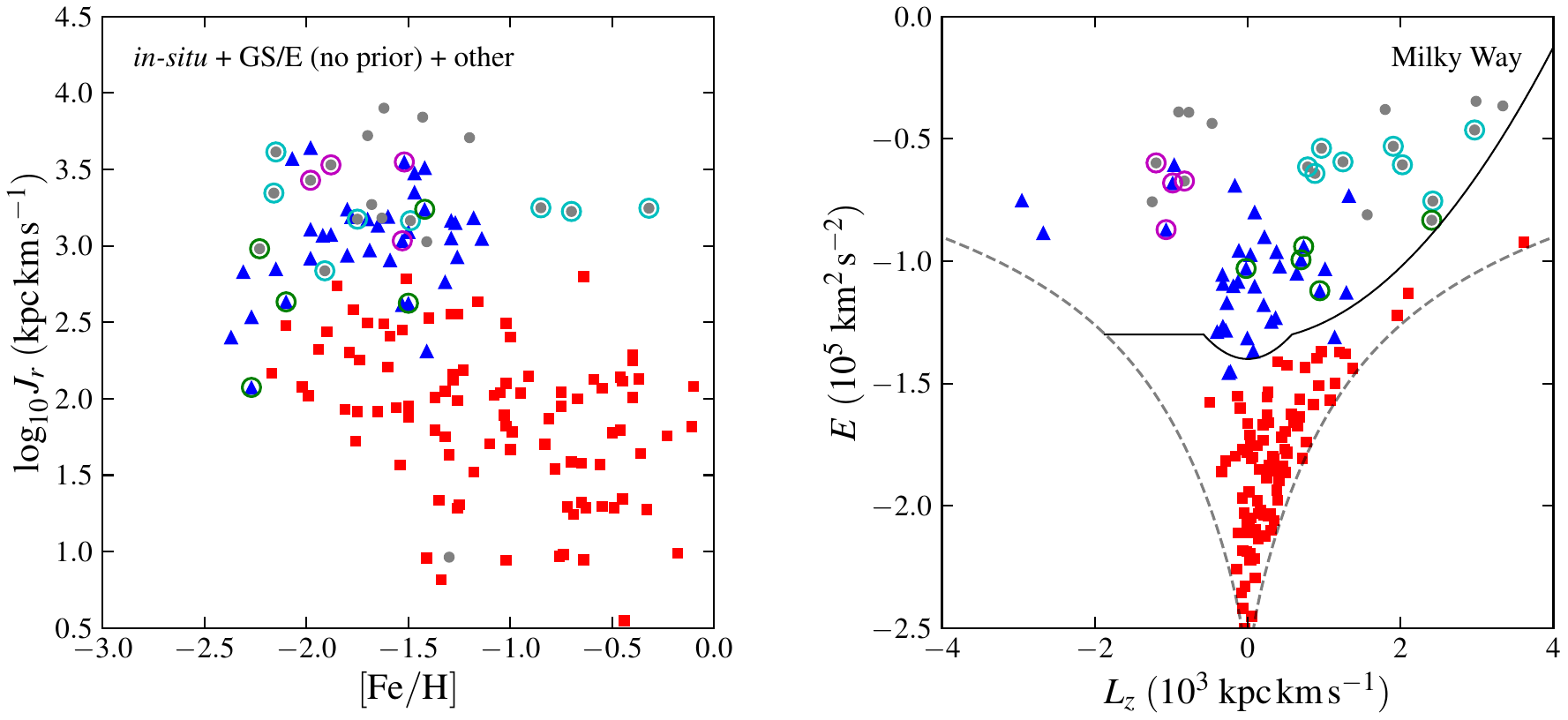} \\
    \vspace{1mm}
    \includegraphics[width=0.8\linewidth]{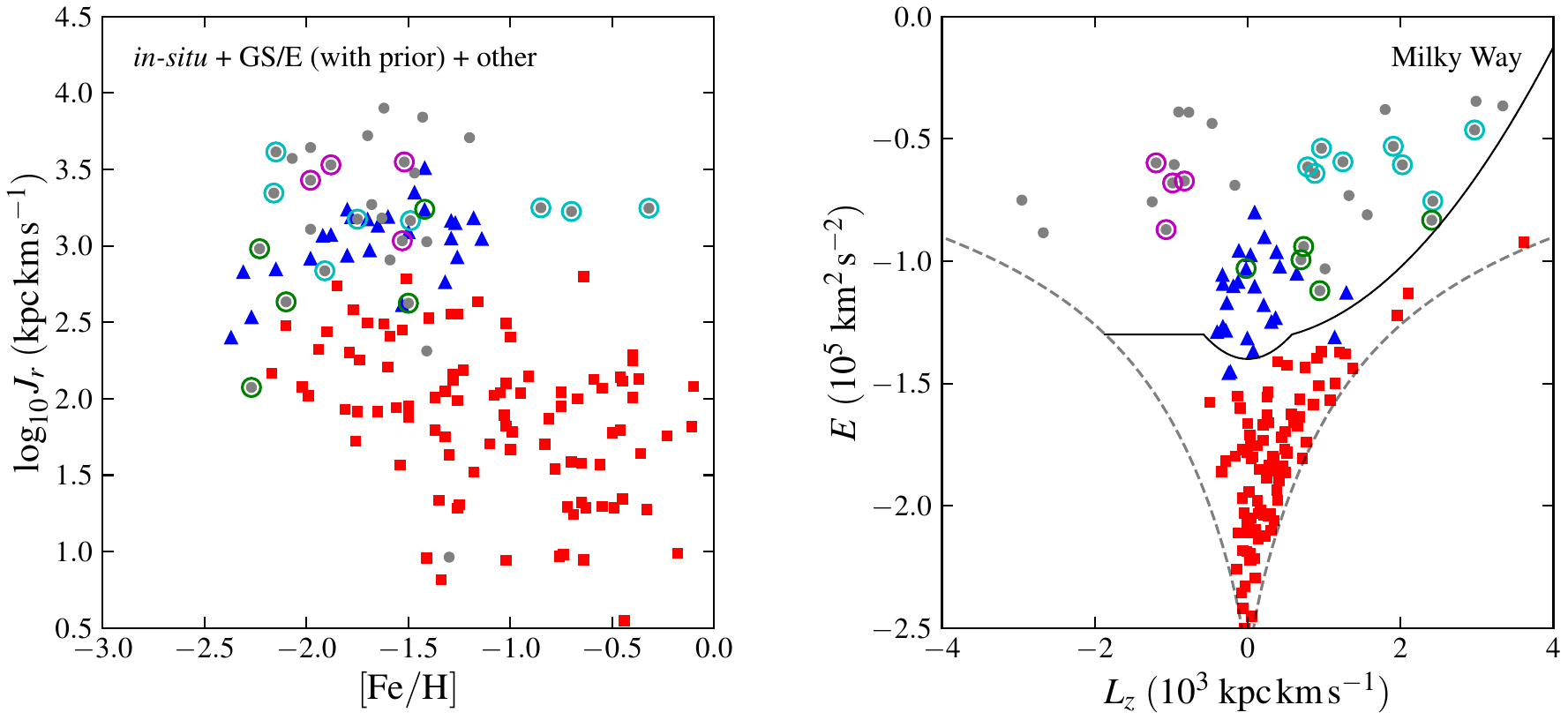} \\
    \vspace{1mm}
    \includegraphics[width=0.8\linewidth]{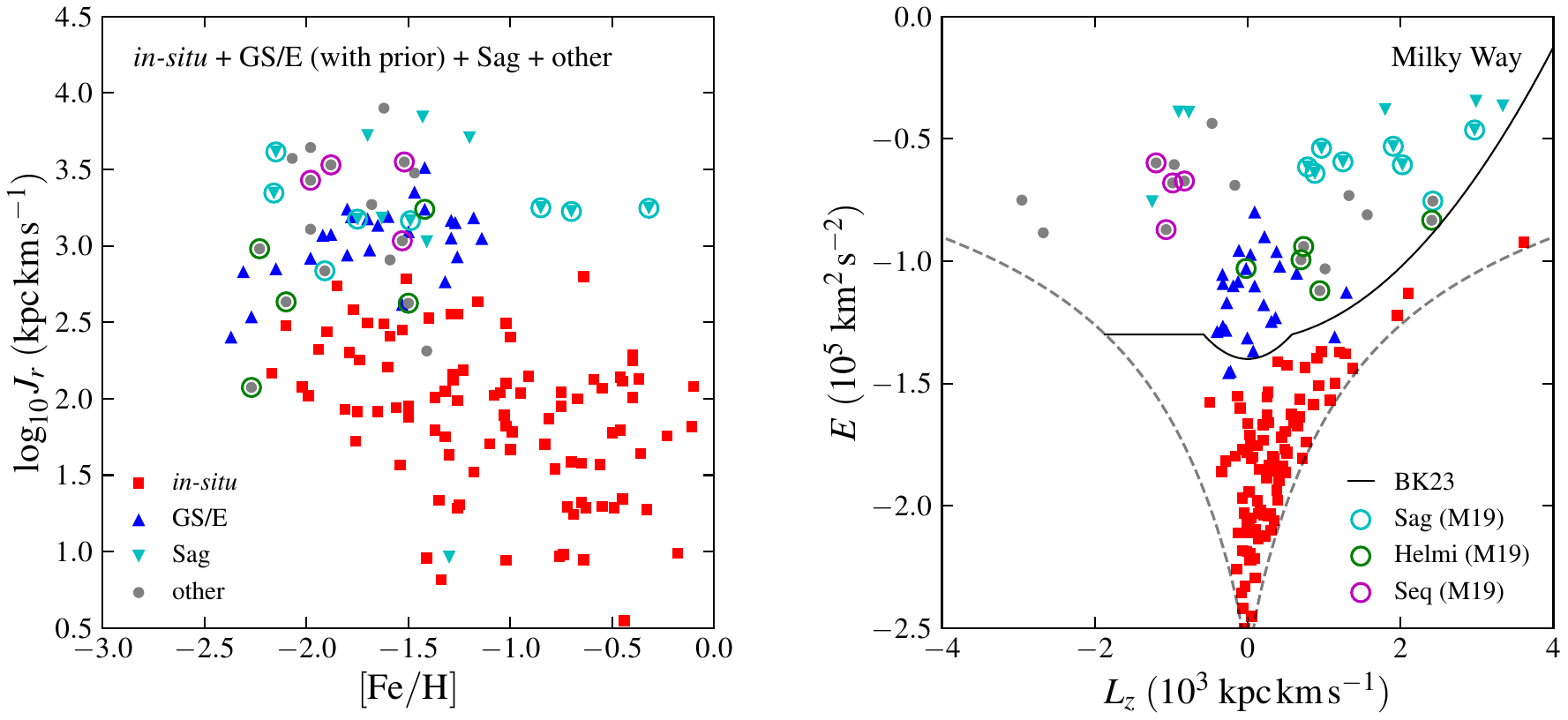} \\
    \vspace{1mm}
    \caption{Classification results for the Galactic GCs in the $\feh$--$\Jr$ space (\textit{left column}) and IoM space (\textit{right column}). The results are combined from various classifiers using a Bayesian approach, as detailed in the text. All panels show the \textit{in-situ} population as red squares. This population is the same for all rows because our classification of \textit{in-situ} vs. \textit{ex-situ} GCs is robust. The \textit{top row} shows the $N_{\rm g}=2$ classification of all \textit{ex-situ} GCs without prior knowledge, i.e., purely unsupervised clustering. We mark the group associated with GS/E as blue triangles. For comparison, we also show the GCs unambiguously assigned to the Sagittarius dwarf (cyan), Helmi streams (green), and Sequoia galaxy (magenta) by \citet[][M19]{massari_origin_2019} as open circles. In the \textit{middle row}, we incorporate prior knowledge by excluding the classifiers that put more than 1 unambiguous GC from any of the three progenitors to GS/E. Based on this classification, we again split the remaining GCs into two groups (\textit{bottom row}). We highlight the group corresponding to the Sagittarius dwarf as cyan triangles. For reference, we plot the circular angular momentum $L_{\rm circ}$ in the IoM space as the dashed curves. We also show the classification boundary by \citet[][BK23]{belokurov_-situ_2024} in the IoM space as the solid curve for comparison.}
    \label{fig:classification_mw}
\end{figure*}

We illustrate the  classification result in Fig.~\ref{fig:classification_mw}. There are 94 out of the 150 Galactic GCs being classified as \textit{in-situ}. The majority of the \textit{in-situ} clusters are located in the high-metallicity and low-$J_r$ region of the $\feh$--$\Jr$ space and the low-energy and positive-$L_z$ region of the IoM space. There are around 10 \textit{in-situ} clusters located in relatively high-energy region with $\E>1.5\times10^5\ {\rm km^2\,s^{-2}}$. However, these clusters are still distinguishable from the \textit{ex-situ} GCs since they are on circular and prograde orbits with $L_z$ close to the maximum value.

It is notable that the posterior probability only depends weakly on the prior $P_{\rm prior}$ in our analysis. We find $1\%<P_{\rm prior}<99\%$ doesn't change the posterior classification of \textit{in-situ} vs. \textit{ex-situ} clusters. For most selections of $P_{\rm prior}$, the posterior probability is extremely close to either 0 or 1 with offset even less than 1\%, supporting the robustness of the classification results.

As mentioned before, there would be more than 100 valid classifiers if we reduce the accuracy threshold to a lower value of 75\%. While an increase in the number of classifiers makes the analysis more comprehensive, having too many classifiers may overestimate the confidence level of the results. This is because a large group of correlated classifiers violates the assumption of independence. When many classifiers are correlated, the effective number of independent classifiers becomes overestimated, leading to over-confidence in the posterior probabilities. Nevertheless, despite this potential issue, the actual results using a threshold of either 78\% or 75\% are very similar, with a difference of only one GC.

For comparison with \citet{belokurov_-situ_2024}, we include their boundary in the IoM spaces on Fig.~\ref{fig:classification_mw}. Based on whether a GC has a lower or higher energy than this boundary, they classified GCs as either \textit{in-situ} or \textit{ex-situ}, respectively. Our classification results largely align with theirs, and only three \textit{ex-situ} clusters in this work (NGC 6205, 7078, and 7099) were assigned as \textit{in-situ} by these authors. These three clusters are located near the boundary in nearly all nine variables we considered, while being slightly more clustered towards the \textit{ex-situ} group in our analysis. The alignment with \citet{belokurov_-situ_2024} supports their conclusion that the boundary in the IoM space provides reliable classification. For the case of the MW we cannot significantly improve the separation by adding 8 more observables. However, this does not indicate that the other observables are useless. We need all the observables to accurately detect the clustering features and to define the hyperplane separating the \textit{in-situ} and \textit{ex-situ} clusters. Our result shows that this hyperplane is near-perpendicular to the IoM space, where we can simplify the separation into a line.

Moreover, all \textit{in-situ} GCs by \citet{massari_origin_2019}, except for one (NGC 7078), are also classified as \textit{in-situ} in this work. Notably, the group they referred to as the ``low-energy group'' is now entirely classified as \textit{in-situ}, in agreement with \citet{belokurov_-situ_2024}, who found that this group has no distinct clustering characteristics in the IoM space or the $\rm[Mg/Fe]$--$\rm[Al/Fe]$ plane. Here, we extend the conclusion to all nine variables, as our clustering methods do not detect distinct clustering signatures for the ``low-energy group'' across all nine GC properties we considered. Conversely, the ``high-energy group'' identified by \citet{massari_origin_2019} is now classified as \textit{ex-situ} in our analysis. 

\begin{table*}
    \caption{Predicted progenitors for 150 MW GCs analyzed in this work (\textit{middle columns}) and \citet[][M19, \textit{right columns}]{massari_origin_2019}. ``L-E'' and ``H-E'' are the low-energy and high-energy groups, respectively; ``Sag'' stands for the Sagittarius dwarf; ``Helmi'' for the Helmi streams; and ``Seq'' for the Sequoia galaxy.\\ A plain ASCII version is available at \url{https://umich.edu/~ognedin/mw_gc_classification.txt}.}
    \label{tab:apply_to_mw}
    \centering
    \renewcommand\arraystretch{1.4} 
    \begin{tabular}{ccc|ccc|ccc}
    \hline\hline
    Cluster & This work & M19 & Cluster & This work & M19 & Cluster & This work & M19 \\
    \hline
    1636-283 & \textit{in-situ} & \textit{in-situ} & Pal 11 & \textit{in-situ} & \textit{in-situ} & NGC 2808 & GS/E & GS/E \\
    BH 261 & \textit{in-situ} & \textit{in-situ} & Ter 1 & \textit{in-situ} & \textit{in-situ} & NGC 288 & GS/E & GS/E \\
    Djor 2 & \textit{in-situ} & \textit{in-situ} & Ter 2 & \textit{in-situ} & \textit{in-situ} & NGC 362 & GS/E & GS/E \\
    HP 1 & \textit{in-situ} & \textit{in-situ} & Ter 3 & \textit{in-situ} & \textit{in-situ} & NGC 4147 & GS/E & GS/E \\
    IC 1276 & \textit{in-situ} & \textit{in-situ} & Ter 4 & \textit{in-situ} & \textit{in-situ} & NGC 5286 & GS/E & GS/E \\
    Lynga 7 & \textit{in-situ} & \textit{in-situ} & Ter 5 & \textit{in-situ} & \textit{in-situ} & NGC 6205 & GS/E & GS/E \\
    NGC 104 & \textit{in-situ} & \textit{in-situ} & Ter 6 & \textit{in-situ} & \textit{in-situ} & NGC 6229 & GS/E & GS/E \\
    NGC 4372 & \textit{in-situ} & \textit{in-situ} & Ter 9 & \textit{in-situ} & \textit{in-situ} & NGC 6341 & GS/E & GS/E \\
    NGC 5927 & \textit{in-situ} & \textit{in-situ} & Ter 12 & \textit{in-situ} & \textit{in-situ} & NGC 6779 & GS/E & GS/E \\
    NGC 6171 & \textit{in-situ} & \textit{in-situ} & NGC 5946 & \textit{in-situ} & L-E & NGC 6864 & GS/E & GS/E \\
    NGC 6218 & \textit{in-situ} & \textit{in-situ} & NGC 5986 & \textit{in-situ} & L-E & NGC 7089 & GS/E & GS/E \\
    NGC 6266 & \textit{in-situ} & \textit{in-situ} & NGC 6093 & \textit{in-situ} & L-E & NGC 7099 & GS/E & GS/E \\
    NGC 6293 & \textit{in-situ} & \textit{in-situ} & NGC 6121 & \textit{in-situ} & L-E & NGC 7492 & GS/E & GS/E \\
    NGC 6304 & \textit{in-situ} & \textit{in-situ} & NGC 6139 & \textit{in-situ} & L-E & Pal 2 & GS/E & GS/E? \\
    NGC 6316 & \textit{in-situ} & \textit{in-situ} & NGC 6144 & \textit{in-situ} & L-E & NGC 5634 & GS/E & Helmi/GS/E \\
    NGC 6325 & \textit{in-situ} & \textit{in-situ} & NGC 6254 & \textit{in-situ} & L-E & NGC 5904 & GS/E & Helmi/GS/E \\
    NGC 6342 & \textit{in-situ} & \textit{in-situ} & NGC 6256 & \textit{in-situ} & L-E & NGC 6981 & GS/E & Helmi \\
    NGC 6352 & \textit{in-situ} & \textit{in-situ} & NGC 6273 & \textit{in-situ} & L-E & NGC 7078 & GS/E & \textit{in-situ} \\
    NGC 6355 & \textit{in-situ} & \textit{in-situ} & NGC 6287 & \textit{in-situ} & L-E & NGC 6426 & GS/E & H-E \\
    NGC 6356 & \textit{in-situ} & \textit{in-situ} & NGC 6333 & \textit{in-situ} & L-E & NGC 6584 & GS/E & H-E \\
    NGC 6362 & \textit{in-situ} & \textit{in-situ} & NGC 6401 & \textit{in-situ} & L-E & Arp 2 & Sag & Sag \\
    NGC 6366 & \textit{in-situ} & \textit{in-situ} & NGC 6402 & \textit{in-situ} & L-E & NGC 2419 & Sag & Sag \\
    NGC 6380 & \textit{in-situ} & \textit{in-situ} & NGC 6441 & \textit{in-situ} & L-E & NGC 6715 & Sag & Sag \\
    NGC 6388 & \textit{in-situ} & \textit{in-situ} & NGC 6453 & \textit{in-situ} & L-E & Pal 12 & Sag & Sag \\
    NGC 6397 & \textit{in-situ} & \textit{in-situ} & NGC 6517 & \textit{in-situ} & L-E & Ter 7 & Sag & Sag \\
    NGC 6440 & \textit{in-situ} & \textit{in-situ} & NGC 6541 & \textit{in-situ} & L-E & Ter 8 & Sag & Sag \\
    NGC 6496 & \textit{in-situ} & \textit{in-situ} & NGC 6544 & \textit{in-situ} & L-E & Whiting 1 & Sag & Sag \\
    NGC 6522 & \textit{in-situ} & \textit{in-situ} & NGC 6681 & \textit{in-situ} & L-E & AM 1 & Sag & H-E \\
    NGC 6528 & \textit{in-situ} & \textit{in-situ} & NGC 6712 & \textit{in-situ} & L-E & Eridanus & Sag & H-E \\
    NGC 6539 & \textit{in-situ} & \textit{in-situ} & NGC 6809 & \textit{in-situ} & L-E & Pal 3 & Sag & H-E \\
    NGC 6540 & \textit{in-situ} & \textit{in-situ} & Pal 6 & \textit{in-situ} & L-E & Pal 4 & Sag & H-E \\
    NGC 6553 & \textit{in-situ} & \textit{in-situ} & Ton 2 & \textit{in-situ} & L-E & Pyxis & Sag & H-E \\
    NGC 6558 & \textit{in-situ} & \textit{in-situ} & NGC 6535 & \textit{in-situ} & L-E/Seq & AM 4 & Sag & - \\
    NGC 6569 & \textit{in-situ} & \textit{in-situ} & Djor 1 & \textit{in-situ} & GS/E & NGC 5694 & other & H-E \\
    NGC 6624 & \textit{in-situ} & \textit{in-situ} & NGC 4833 & \textit{in-situ} & GS/E & NGC 6934 & other & H-E \\
    NGC 6626 & \textit{in-situ} & \textit{in-situ} & NGC 5897 & \textit{in-situ} & GS/E & Pal 14 & other & H-E \\
    NGC 6637 & \textit{in-situ} & \textit{in-situ} & NGC 6235 & \textit{in-situ} & GS/E & NGC 5824 & other & Sag \\
    NGC 6638 & \textit{in-situ} & \textit{in-situ} & NGC 6284 & \textit{in-situ} & GS/E & Pal 15 & other & GS/E? \\
    NGC 6642 & \textit{in-situ} & \textit{in-situ} & Ter 10 & \textit{in-situ} & GS/E & NGC 6101 & other & Seq/GS/E \\
    NGC 6652 & \textit{in-situ} & \textit{in-situ} & NGC 5139 & \textit{in-situ} & GS/E/Seq & NGC 3201 & other & Seq/GS/E \\
    NGC 6656 & \textit{in-situ} & \textit{in-situ} & E 3 & \textit{in-situ} & Helmi? & IC 4499 & other & Seq \\
    NGC 6717 & \textit{in-situ} & \textit{in-situ} & 2MASS-GC02 & \textit{in-situ} & - & NGC 5466 & other & Seq \\
    NGC 6723 & \textit{in-situ} & \textit{in-situ} & Liller 1 & \textit{in-situ} & - & NGC 7006 & other & Seq \\
    NGC 6749 & \textit{in-situ} & \textit{in-situ} & UKS 1 & \textit{in-situ} & - & Pal 13 & other & Seq \\
    NGC 6752 & \textit{in-situ} & \textit{in-situ} & ESO-SC06 & GS/E & GS/E & NGC 4590 & other & Helmi \\
    NGC 6760 & \textit{in-situ} & \textit{in-situ} & IC 1257 & GS/E & GS/E & NGC 5024 & other & Helmi \\
    NGC 6838 & \textit{in-situ} & \textit{in-situ} & NGC 1261 & GS/E & GS/E & NGC 5053 & other & Helmi \\
    Pal 1 & \textit{in-situ} & \textit{in-situ} & NGC 1851 & GS/E & GS/E & NGC 5272 & other & Helmi \\
    Pal 8 & \textit{in-situ} & \textit{in-situ} & NGC 1904 & GS/E & GS/E & Pal 5 & other & Helmi? \\
    Pal 10 & \textit{in-situ} & \textit{in-situ} & NGC 2298 & GS/E & GS/E & Rup 106 & other & Helmi? \\
    \hline
    \hline
    \end{tabular}
    \vspace{2mm}
\end{table*}

Next, we focus on the \textit{ex-situ} clusters and split them into two groups. One group stands for the GS/E and the other for the remaining clusters. We use a Bayesian approach adopting classifiers' accuracy evaluated from \S\ref{sec:n_g_2}. The group with lower energy is labeled GS/E. We lower the accuracy threshold from 78\% to 69\% to obtain a sufficient number of valid classifiers since the accuracy of identifying the GS/E population is generally lower. We show this classification in the \textit{top row} of Fig.~\ref{fig:classification_mw}.

Our classification associates 37 \textit{ex-situ} GCs to the GS/E merger. These GCs are in the low-energy region and have lower $J_r$ but similar metallicity to other \textit{ex-situ} GCs. This is mostly consistent with observations, except for the Helmi streams and the Sequoia galaxy. For example, \citet{massari_origin_2019} unambiguously assigned five GCs to the Helmi streams and four to the Sequoia galaxy. However, seven of them are categorized to the GS/E in our classification, indicating that these GCs have indistinguishable clustering features to the GCs from the GS/E. 

We state that ``unambiguous GCs'' refer to GCs assigned to a single progenitor by \citet{massari_origin_2019}. It is not necessarily true that such assignments are absolutely correct. However, in order to align more closely with the prior knowledge and produce classification more consistent with generally agreed results, we further refine our approach by excluding classifiers that assign more than one GC from any one of the unambiguous groups (Sagittarius, Helmi, and Sequoia) by \citet{massari_origin_2019}. The updated classification, shown in the \textit{middle row} of Fig.~\ref{fig:classification_mw}, associates 26 GCs with the GS/E. Compared to the previous ``no prior'' classification, only one unambiguous cluster that was assigned to the Helmi streams by \citet{massari_origin_2019}, NGC 6981, is still in the GS/E group. NGC 6981 was also associated with the ``Kraken'' merger by \citet{kruijssen_formation_2019}. This uncertainty suggests that more data beyond the variables we used are needed to accurately determine this cluster's progenitor.

Interestingly, only two clusters in the ``high-energy group'' by \citet{massari_origin_2019} are associated with the GS/E in this work, while the rest are linked to other progenitor galaxies, supporting their interpretation that the ``high-energy group'' consists of clusters accreted from various low-mass progenitors. 

Finally, we aim to identify the second most massive merger from the \textit{ex-situ} GCs excluding the GS/E. Direct classification into three groups ($N_{\rm g}=3$ case in \S\ref{sec:n_g_greater_than_3}) does not yield any result that can be self-consistently explained by prior knowledge. We thus do not present results from this approach. We also do not proceed to identify more mergers since none of the classifiers show convincing performance for more than three groups (\S\ref{sec:n_g_greater_than_3}). Since the classification of GS/E is refined with prior knowledge and is more likely to be correct, we proceed with the recursive approach in \S\ref{sec:n_g_greater_than_3} to split remaining GCs into two groups, hoping that the error of classifying GS/E is less influential. 

Using the Bayesian method with an accuracy threshold of 68\%, we find that one of the groups resembles the Sagittarius dwarf. This group consists of 13 GCs, located in the high-energy and \(L_z \gtrsim 0\) region in the IoM space. Remarkably, seven of the eight GCs unambiguously assigned to the Sagittarius dwarf by \citet{massari_origin_2019} fall into this group. The one exception is NGC 5824, which was also not among the six most definite Sagittarius GCs suggested by these authors. The exception is consistent with \citet{bellazzini_globular_2020}, where NGC 5824 fails to meet all three selection criteria by these authors. No GC from other unambiguous progenitors falls into this group. We detail the classification results for individual GCs in Tab.~\ref{tab:apply_to_mw}, along with the comparison with \citet{massari_origin_2019}.

\section{Discussion}
\label{sec:discussion}

\subsection{Evolution of orbital parameters}
\label{sec:evolution}

The idea of identifying progenitor galaxies using GC properties is built upon the assumption that GCs preserve intrinsic properties they inherited from their original host galaxies. These properties are presumed to be distinct enough from those of GCs of other populations. In this subsection, we investigate the validity of this assumption, exploring how galaxy mergers influence the distribution of GC orbital properties. 

We use galaxy \texttt{519311} as a case study. This galaxy has a notable number of significant mergers (five in total, considering the \textit{early-accreted} one) and has been well studied in previous works \citep[e.g.][]{chandra_three-phase_2023}. The insights and conclusions from this case study apply to other galaxies as well. This investigation helps us understand the degree to which GCs maintain their original characteristics during key galaxy assembly events. 

We study the distribution of GCs of \texttt{519311} in various property spaces to understand how their properties evolved over time. In Fig.~\ref{fig:evolve} we present the evolution of GC distributions in the IoM ($L_z$--$E$), normalized IoM ($\varepsilon$--$e$), and action ($J_r$--$J_z$) spaces. When computing the energy and actions at a past epoch, we define the orientation of the galaxy disk based on the angular momentum of all stellar particles in the galaxy at that time. Note that this orientation was not static. It has rotated by approximately 60 degrees over time, as detailed in \citet{chandra_three-phase_2023}. For clarity, in every panel we show the same GCs that survived to the present. 

We focus specifically on four important epochs: 1) before the merger with the \textit{early-accreted} galaxy \texttt{519311-c}, 2) after the merger with \texttt{c}, 3) before the merger with the most dominant satellite \texttt{519311-b}, and 4) after the merger with \texttt{b}. We discuss in detail these two significant mergers in the following subsections.

\subsubsection{Impact of the \textit{early-accreted} galaxy}

The satellite galaxy \texttt{519311-c} is an ancient galaxy that merged with the central galaxy more than 11.4 billion years ago. It quickly dissolved within 1~Gyr after the infall. As shown in Fig.~\ref{fig:mh_log_vs_tlookback}, \texttt{519311-c} was once a galaxy of comparable mass to the central galaxy. Therefore, the merger was quite violent allowing a significant fraction of the merging GCs to penetrate the low-energy region near the galactic center. These GCs also tend to have lower $J_r$ and $J_z$ as they migrated to inner orbits within the central galaxy.

The merger not only affected the GCs from \texttt{519311-c} itself but also had a notable impact on the \textit{in-situ} clusters. The normalized energy of the \textit{in-situ} clusters increased from $e\approx -0.7$ to $-0.6$ as a result of the merger. The normalized energy of the \textit{in-situ} clusters remained relatively stable for the next 11~Gyr after the merger with \texttt{519311-c}.

The merger with the \textit{early-accreted} satellite strongly influenced the distributions of both merging and \textit{in-situ} GCs. Consequently, the two GC populations were significantly mixed in all three spaces we examined. Such extensive mixing makes it challenging to differentiate these GCs based solely on their kinematics. Furthermore, both the GCs from the satellite and the old \textit{in-situ} population (age $>11.4$~Gyr) have comparable metallicities. This is because the host galaxies of both populations had similar mass at the time of GC formation. In our model a cluster's metallicity is determined by the formation time and the mass of its host galaxy, see Eq.~(6) in \citetalias{chen_catalogue_2024}.

Consequently, it is challenging to distinguish ancient mergers with galaxies of comparable masses to the central from the \textit{in-situ} component. Moreover, the concept of a ``main-progenitor galaxy'' is less clear during the early stages of galaxy formation. Many ``satellites'' during this period were almost as massive as the main progenitor, leading to a ``bottom-up'' assembly where small galaxies combined to form larger structures. Given these considerations, we choose not to differentiate between the \textit{early-accreted} GCs and the \textit{in-situ} ones as mentioned in \S\ref{sec:in_vs_ex}.

\subsubsection{Impact of the most dominant merger}

\begin{figure*}
    \centering
    \includegraphics[width=\linewidth]{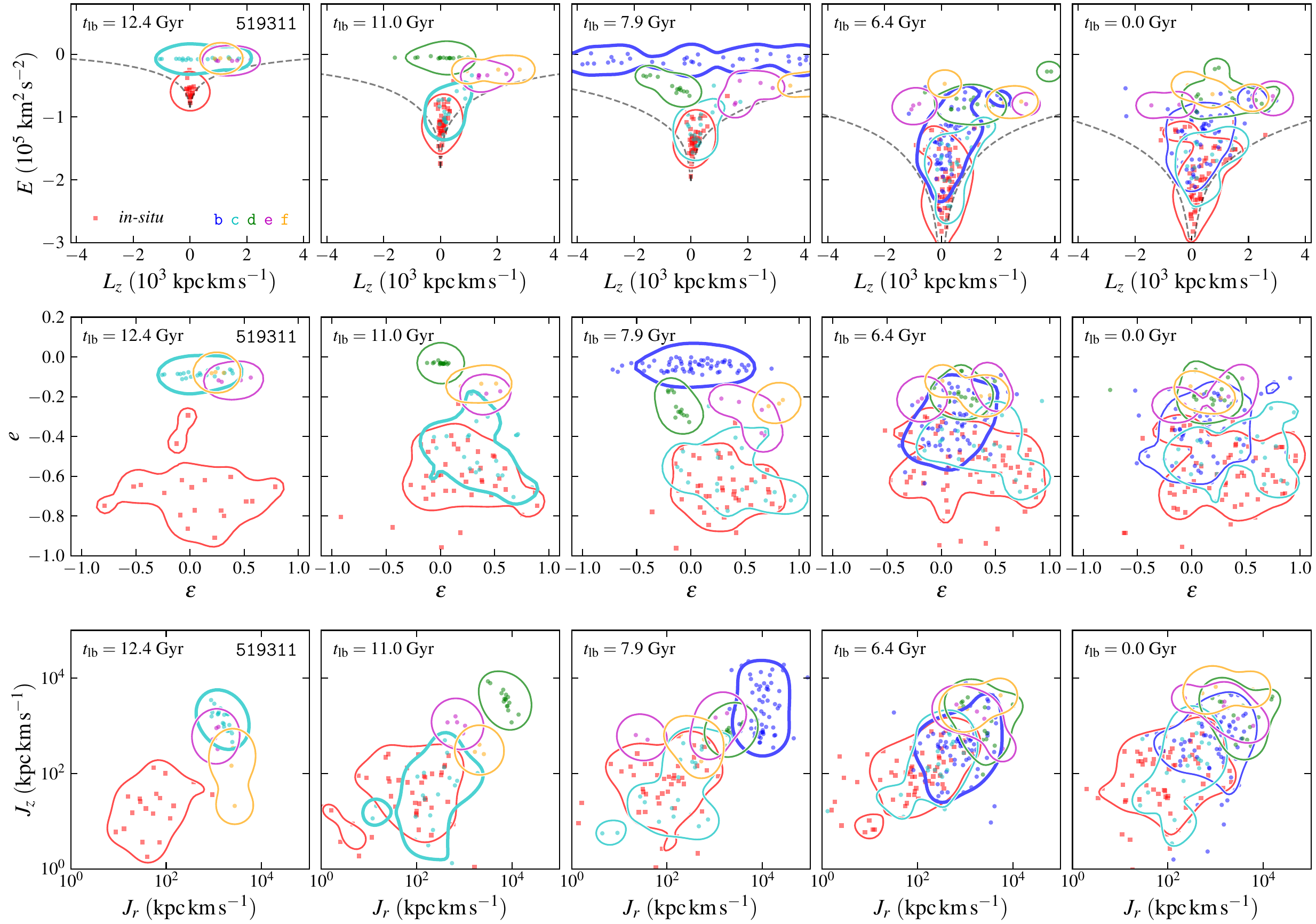}
    \vspace{-1mm}
    \caption{Evolution of GCs that survive to the present day in galaxy \texttt{519311} in the IoM space ($L_z$--$E$, \textit{top row}), normalized IoM space ($\varepsilon$--$e$, \textit{middle row}), and action space ($J_r$--$J_z$, \textit{bottom row}). The contours enclose 75\% of GC number (after Gaussian KDE smoothing) in each progenitor. From left to right, the 5 columns refer to the epochs 1) before the merger with \texttt{519311-c}, 2) after the merger with \texttt{c}, 3) before the merger with \texttt{519311-b}, 4) after the merger with \texttt{b}, and 5) at present. We emphasize the merging \texttt{519311-c} and \texttt{b} GCs as thicker contours in the relevant panels.}
    \label{fig:evolve}
\end{figure*}

\begin{figure*}
    \centering
    \includegraphics[width=0.9\linewidth]{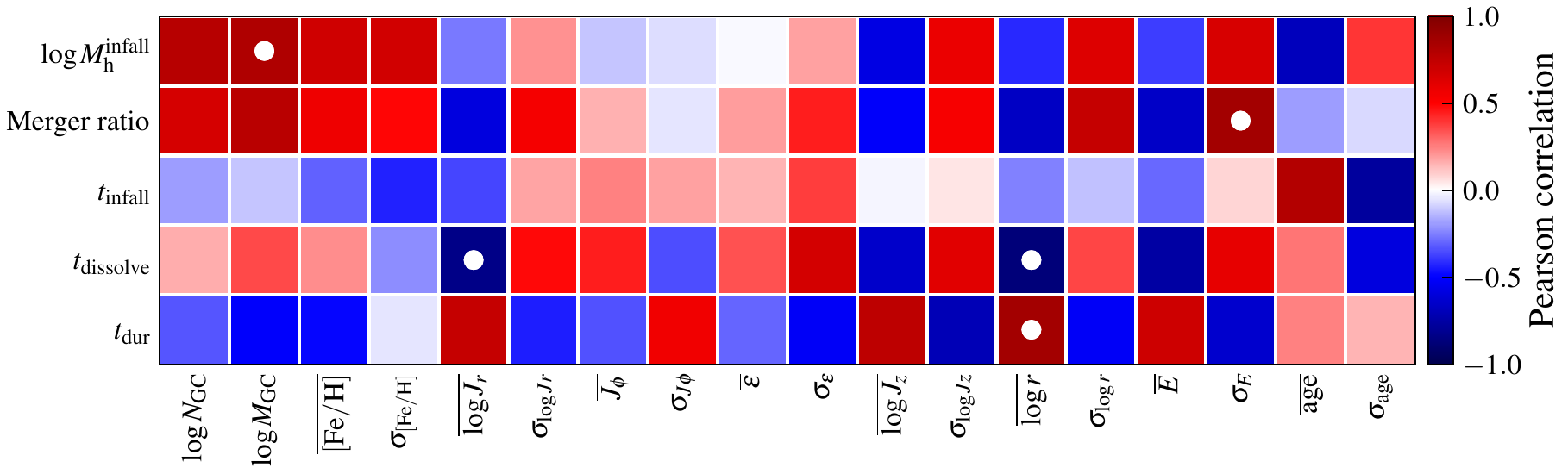}
    \vspace{0mm}
    \caption{Pearson correlation map between properties of progenitor galaxies and the surviving GC populations. Important pairs with the correlation coefficient $|r|>0.8$ are highlighted with white circles.}
    \label{fig:cov_map_merger_vs_gc}
\end{figure*}

The merger between \texttt{519311} and \texttt{519311-b} is a key event in this galaxy's assembly history. This merger resembles the GS/E merger in several aspects. It began around $\tlb=8$~Gyr and continued for $2-3$~Gyr until the satellite fully dissolved. The orbit was almost radial, and the mass ratio was close to unity.

Although the merger ratio is high, the GCs from \texttt{b} did not penetrate the low-energy region near the galactic center as deeply as those from \texttt{519311-c}. This is likely because 1) the merger with \texttt{b} lasted much longer than the merger with \texttt{c}, resulting in a less violent process compared to the intense merger with \texttt{c}; and 2) the central galaxy already developed a deeper gravitational potential well at the time of merger with \texttt{b}, preventing the merging GCs from \texttt{b} from reaching the inner orbits of low energy.

However, GCs from \texttt{b} still reached lower-energy regions than other mergers with \texttt{d}, \texttt{e}, and \texttt{f}. These smaller satellites were accreted and dissolved before \texttt{b} (but still after \texttt{c}). We notice that the GCs of these less dominant mergers had more clustered and distinct distributions in the IoM space and action space prior to the accretion of \texttt{b}. However, the subsequent merger with \texttt{b} broadened the distribution of actions of these GCs and caused significant overlap.

To understand how much the merger with \texttt{b} made the GCs from \texttt{d}, \texttt{e}, and \texttt{f} less distinct from the rest of the GCs, we need to quantify the distinction of satellite GC populations before and after the merger. This can be measured using the KL divergence between the GCs from a given satellite and all other GCs. The KL divergence \citep{kullback_information_1951} is a statistical measure quantifying the distance between two probability density functions (PDFs), say, $p$ and $q$. It is defined as:
\begin{equation}
    D_{\rm KL}(p\|q)\equiv \int_{-\infty}^{\infty} p(x)\ln\frac{p(x)}{q(x)}dx.
    \label{eq:KL}
\end{equation}
If $p$ and $q$ are identical, we have $D_{\rm KL}=0$, indicating that the two PDFs are indistinguishable. The greater the KL divergence, the easier it is to separate the two PDFs. For reference, the KL divergence between two 1D univariate Gaussian distributions with the difference of the means equal 2 standard deviations is $D_{\rm KL}=2$. These two distributions have an overlap fraction of 32\%, which can serve as an estimate of expected classification error. The divergence $D_{\rm KL}=5$ corresponds to a displacement $=\sqrt{10}$ standard deviations, with an overlap fraction of 11\%.

We set $p$ as the PDF of GCs from a given progenitor, and $q$ as the PDF of all other GCs, including the in-situ ones. To calculate the probability distribution from discrete data points, we apply the Gaussian kernel density estimation (KDE) technique. In the IoM space, we set the bandwidth of the kernel to be $\Delta L_z=0.4\times10^3\ {\rm kpc\,km\,s^{-1}}$ and $\Delta\E=0.15\times10^5\ {\rm km^2\,s^{-2}}$. These values are selected to approximate the average separation between GCs in the $L_z$ and $\E$ directions, respectively. Similarly, we set bandwidth to be $\Delta\varepsilon=0.12$, $\Delta e=0.06$, and $\Delta\logJr=\Delta\logJz=0.3$~dex for the normalized IoM space and action space. We plot the contours enclosing $75\%$ of the total number of GCs for each progenitor in Fig.~\ref{fig:evolve}. Then, we numerically approximate the KL divergence utilizing importance sampling:
\begin{equation*}
    D_{\rm KL}(p\|q)\approx \frac{1}{n}\sum_{x\in X} \frac{p(x)}{r(x)}\ln\frac{p(x)}{q(x)}
\end{equation*}
where $X$ is a set of $n$ points sampled from distribution $r$. To simplify this for our case, we take $X$ to be the GCs from the progenitor, which by construction follow the distribution $p$. Thus, the expression becomes
\begin{equation*}
    D_{\rm KL}(p\|q)\approx \frac{1}{n}\sum_{x\in X} \ln\frac{p(x)}{q(x)}.
\end{equation*}

Before the merger with \texttt{519311-b}, the KL divergence values for \texttt{d}, \texttt{e}, and \texttt{f} in the IoM space were 5.1, 4.1, and 5.4, respectively. Given the reference values for the Gaussian distributions, that should be sufficient to classify the satellite GCs with an accuracy $\ga90\%$. After the merger, these values decreased to 2.8, 3.5, and 2.9. That is a significant change that makes the classification more difficult. We observe similar trends in the normalized IoM space, with KL divergence decreasing from 4.7 to 3.3 (\texttt{d}), 3.3 to 2.3 (\texttt{e}), and 5.1 to 1.8 (\texttt{f}); and in the action space: 2.2 to 1.1 (\texttt{d}), 2.3 to 1.5 (\texttt{e}), and 2.4 to 1.8 (\texttt{f}). These numbers quantitatively demonstrate how the GS/E-like merger disturbed the GC distributions from previous satellites, making them less distinguishable. The impact of the merger was significant due to its mass and the reorientation of the galactic disk by about 60 degrees, which altered the computation of orbital actions, especially the orientation-sensitive $L_z$.

After the merger with \texttt{519311-b}, the IoM and action distributions for GCs from all progenitors remained relatively stable over the next 6~Gyr, up to the present. The KL divergence for \texttt{d}, \texttt{e}, and \texttt{f} showed little change, with some even increased slightly up to 30\%. This stability suggests that energy and actions are reliable tracers of the progenitor's properties in the absence of major disruptive merger events. This conclusion is valid in the last 10~Gyr of the MW assembly history, which is relatively quiescent. However, the significant perturbations by GS/E had a significant effect on the kinematics of GCs accreted before the merger, making them less distinguishable. This disruptive merger also explains the difficulty in identifying smaller progenitor galaxies beyond GS/E, as discussed in \S\ref{sec:n_g_greater_than_3}. A real-world example of such progenitors influenced by the GS/E is the Sequoia event, which likely merged with the Galaxy around the same time as or prior to the GS/E \citep{myeong_evidence_2019,forbes_reverse_2020,valenzuela_galaxy_2023}. Conversely, the Sagittarius dwarf galaxy merged later, preventing its GC kinematics from being perturbed by the GS/E. Consequently, independent studies have consistently classified the GCs from Sagittarius as a distinct high-energy group with prograde motion in the IoM space.

\subsection{Merger properties revealed by GCs}
\label{sec:merger_properties}

This section investigates how well GCs reveal the merger characteristics of their progenitor galaxies. We analyze five key features: the infall time $t_{\rm infall}$ when the progenitor reached its maximum mass, the dissolution time $t_{\rm dissolve}$ when the halo finder no longer identifies this galaxy, the merger duration $t_{\rm dur}\equiv t_{\rm infall}-t_{\rm dissolve}$, the progenitor mass at infall $\Mh^{\rm infall}$, and the ratio of $\Mh^{\rm infall}$ to the central galaxy mass at infall. In Fig.~\ref{fig:cov_map_merger_vs_gc} we show the Pearson correlation coefficients $r$ between these merger characteristics and important GC features, including the total number $\Ngc$ and mass $\Mgc$ of surviving clusters, and the mean and standard deviation of the other properties considered in this work.

We note strong correlations ($r>0.8$) between the current radius and apocenter/pericenter radii of GCs. For clarity, we only analyze the current radius in this section. The apocenter/pericenter radii follow similar trend as the radius.

The following subsections discuss the best indicators of these five merger characteristics, focusing on those with strongest correlation or anti-correlation.

\subsubsection{Infall mass}

The total mass of the GC system has the strongest correlation with the mass of their progenitor galaxy at infall. This relationship is remarkably linear, $\Mgc \approx 3\times10^{-5}\Mh^{\rm infall}$. Even more surprising, it is described by the same equation as the well-established relation for present-day galaxies in a very wide range of mass $\Mh=10^{8} - 10^{14}\Msun$ \citep{harris_dark_2015,forbes_extending_2018}: $\Mgc=3\times10^{-5}\Mh$. We show this linear relation in Fig.~\ref{fig:Mgcs_Mh}. We also show the best-fitting power-law relation from \cite{chen_formation_2023}, which is similar: $\Mgc\propto\Mh^{0.93}$. Here we find that the linear relation extends to the merged progenitor galaxies and their surviving GC populations. This result provides a unique proxy of the maximum mass of satellite galaxies at infall based on the mass of their accreted GC population.

The prediction that the $\Mgc$--$\Mh^{\rm infall}$ relation of the progenitor galaxies follows the same scaling as $\Mgc$--$\Mh$ of the surviving galaxies is remarkable. Our model provides two key insights to understand this. First, our model links the initial mass of a cluster population $\Mgc^{\rm init}$ to the gas mass of the progenitor galaxy. At high redshift, when majority of satellite GCs form, the gas mass is nearly proportional to the total halo mass. If each population maintains the linear $\Mgc$--$\Mh$ relation, the sum of all GC populations will also obey the same linear relation. Therefore, the sum of initial masses of GCs in a given progenitor galaxy would be proportional to the galaxy mass at the epoch when GC formation stops. In our model, this happens when the progenitor mass no longer increases by a substantial amount, see \S\ref{sec:model}. This epoch usually corresponds to the infall onto the central galaxy. Thus we expect $\Mgc^{\rm init} \propto \Mh^{\rm infall}$. Fig.~\ref{fig:Mgcs_Mh} shows indeed that satellite galaxies obey such a linear relation, with the normalization $\Mgc^{\rm init} \approx 5\times10^{-4}\Mh^{\rm infall}$. Individual progenitors of course have significant scatter around the mean relation, as a result of the different mass assembly histories leading to the same $\Mh^{\rm infall}$. This ``initial'' GC mass is not an actual mass of a GC system at any epoch because clusters continuously lose mass due to stellar evolution and tidal disruption. The evolution is very significant: by the present time, that initial mass is reduced by a factor of 10$-$20.

Second, as suggested by detailed galaxy formation simulations discussed in \citet{meng_tidal_2022}, the average rate of tidal disruption experienced by \textit{ex-situ} clusters before their progenitor dissolution is similar to that experienced by \textit{in-situ} clusters. Therefore, we may expect comparable reduction of \textit{in-situ} and \textit{ex-situ} $\Mgc$ before infall. After infall, the \textit{ex-situ} clusters are typically located at larger distances from the central galaxy than their \textit{in-situ} counterparts and experience weaker tidal fields. Nevertheless, \citet{meng_tidal_2022} find that at least at $z>1.5$, the \textit{ex-situ} disruption rate is lower only by a factor $1.5-2$. Such a difference is small compared to the 0.4~dex scatter of the $\Mgc$--$\Mh$ relation \citep{chen_formation_2023}. In addition, individual GCs in satellite progenitors typically have lower mass and disrupt faster than more massive \textit{in-situ} clusters for a given strength of the tidal field \citepalias[see \S2.2.2 in][]{chen_catalogue_2024}. The two effects go in the opposite directions and largely cancel each other. Therefore, the dynamical disruption of the \textit{in-situ} and \textit{ex-situ} GCs is similar and preserves the linearity of the $\Mgc$--$\Mh^{\rm infall}$ relation.

\subsubsection{Merger ratio}

The merger ratio, defined as the mass ratio between a progenitor galaxy and the central galaxy at infall, also shows a positive correlation with the total mass of globular clusters, although with a lower correlation coefficient. This is likely a consequence of the $\Mgc$--$\Mh^{\rm infall}$ correlation, with the variation in the central galaxy mass increasing the scatter.

We also notice that the merger ratio correlates with the standard deviation of the orbital energy. This correlation is observed not just for $\E$ but also for most kinematic variables, with the exception of $\Jp$. This is because 1) more massive progenitor galaxies tend to have a greater intrinsic dispersion in the kinematics, and 2) more massive mergers have a more disruptive effect compared to minor ones. They can more effectively disturb the distribution of GCs across the multi-dimensional kinematic space, leading to an increased dispersion in these properties.

\subsubsection{Infall time}

The infall time of a progenitor galaxy is closely linked to the age of its GC population. As the formation of GCs in these progenitor galaxies was shut down after $t_{\rm infall}$, all these clusters must have $t_{\rm infall}< \mathrm{age}<t_{\rm universe}$. Therefore, the mean age increases with $t_{\rm infall}$ while the standard deviation $\sigma_{\rm age}$ decreases with $t_{\rm infall}$. On the other hand, the infall time does not show significant correlation with the other GC properties.

In practice, the age measurement for Galactic GCs typically has errors $\gtrsim 1$~Gyr, which is insufficient to resolve the infall order of the progenitor galaxies, especially since all satellites were likely accreted over a period $\sim$~a few Gyr, as shown in Fig.~\ref{fig:mh_log_vs_tlookback}. However, since calculating the age dispersion requires only relative ages, it is potentially a more reliable indicator of $t_{\rm infall}$ than the average age which requires determination of the absolute ages.

\subsubsection{Dissolution time and merger duration}

The dissolution time $t_{\rm dissolve}$ approximates when satellite GCs were migrated into the central galaxy. This variable has a notable anti-correlation with the mean of $\logr$ and $\logJr$ because the central galaxy was less massive if the progenitor galaxy dissolved earlier, resulting in a shallower gravitational potential. Those GCs could more easily acquire inner low-energy orbits, leading to lower $r$ and $\Jr$.

\begin{figure}
    \centering
    \includegraphics[width=\linewidth]{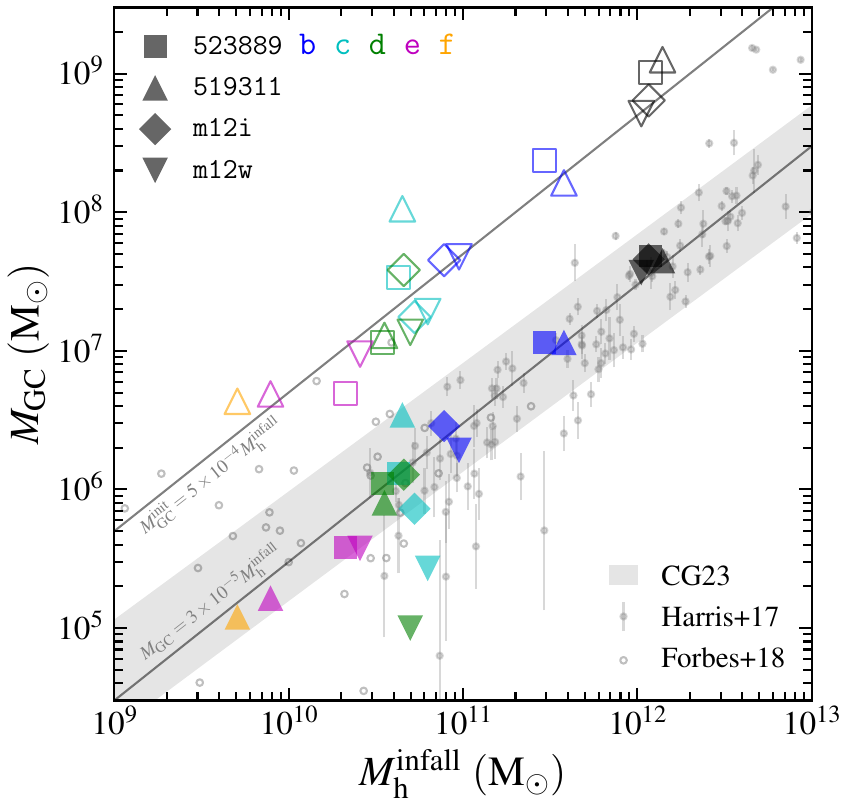}
    \vspace{-3mm}
    \caption{Combined mass of the GC system vs. maximum mass of its progenitor galaxy at infall. The final surviving GC population mass for the progenitor galaxies is shown by filled symbols (color scheme as in Fig.~\ref{fig:mh_log_vs_tlookback}). A linear relation $\Mgc=3\times10^{-5}\Mh^{\rm infall}$ is shown by the lower thin line. We also plot the combined mass of initially formed GCs as open symbols, which can be approximately described by $\Mgc^{\rm init}=5\times10^{-4}\Mh^{\rm infall}$. For comparison, we plot the combined mass of all surviving GCs in the central galaxy against its present-day mass $\Mh$ as the dark gray symbols. In addition, we show the observed $\Mgc$--$\Mh$ relations from \citet[][solid circles with errorbars]{harris_galactic_2017} and \citet[][open circles]{forbes_extending_2018}. These data can be fitted by the power-law function $\Mgc\propto\Mh^{0.93}$ from \citet{chen_formation_2023}, with intrinsic scatter shown as the shaded band.}
    \label{fig:Mgcs_Mh}
\end{figure}

The merger duration $t_{\rm dur}$, defined as $t_{\rm infall}-t_{\rm dissolve}$, is negatively correlated with \(t_{\rm dissolve}\) due to the narrower range of $t_{\rm infall}$ (see Fig.~\ref{fig:mh_log_vs_tlookback}). This means that $t_{\rm dur}$ also carries the opposite correlations of $t_{\rm dissolve}$ with the other GC properties, including mean $\logr$ and $\logJr$. 

\subsection{Comparison with Trujillo-Gomez et al. (2023)}
\label{sec:T23}

The recent study by \citet{trujillo-gomez_situ_2023} has a similar motivation to ours. These authors employed a supervised artificial neural network classifier, trained on around 700 simulated galaxies of various mass from the E-MOSAICS project. The classifier categorizes each GC as \textit{in-situ} or \textit{ex-situ} using only extragalactic observables. By evaluating their method on a test set of around 200 galaxies, they achieved an accuracy $\lesssim90\%$ for the \textit{unambiguously} identified clusters.

Their classifier utilizes a range of input variables from three categories, 1) galaxy properties: total stellar mass, effective radius, mean iron abundance, mean oxygen-to-iron ratio, stellar velocity dispersion; 2) GC system properties: total number of clusters, cluster velocity dispersion; and 3) properties of individual GCs: iron abundance, oxygen-to-iron ratio, projected galactocentric radius, line-of-sight velocity, projected rotation velocity, projected angular momentum. Some of these variables are measured in different units and count as multiple entries. 

The authors also calculated permutation importance, i.e., how much the classification result is affected by randomly shuffling one of the input variables, to assess the impact of each variable on the classification outcome. They found the metallicity (iron abundance and oxygen-to-iron ratio) to be the most important GC property for their purpose. This agrees with our findings, where the metallicity appears as a critical variable in the best configurations for identifying \textit{in-situ} vs. \textit{ex-situ} GCs (see Tables~\ref{tab:configurations} and \ref{tab:configurations_pca}).

Our work differs from \citet{trujillo-gomez_situ_2023} as we specifically focus on MW-analog galaxies and aim to classify not only the \textit{in-situ} vs. \textit{ex-situ} but also the progenitors of \textit{ex-situ} clusters. We therefore analyze a selected sample of simulated galaxies matching various observational constraints on the MW. We also take advantage of the available 3D positions and 3D velocities of MW GCs. This allows us to compute more kinematic characteristics in addition to the extragalactic observables. Therefore, although seemingly similar, the motivation and the application areas of the two works are different.

The inclusion of more kinematic characteristics in our analysis directly improves the classification accuracy. Although \citet{trujillo-gomez_situ_2023} reported an accuracy of around 90\% for the unambiguously classified GCs, a significant number of other clusters remain unclassified. They considered a GC to be unambiguously classified if its probability of being either \textit{in-situ} or \textit{ex-situ} exceeds a threshold of $P_{\rm thresh}=0.79$. This leads to about 40\% of the clusters left as ambiguous. By lowering $P_{\rm thresh}$ to 0.5, where all clusters are classified as either \textit{in-situ} or \textit{ex-situ} (the same as our definition), their accuracy drops to around 80\%. In comparison, our best algorithms achieve $\approx 90\%$ accuracy in categorizing every GC as either \textit{in-situ} or \textit{ex-situ} formed (\S\ref{sec:in_vs_ex}).

In summary, although the classifier developed by \citet{trujillo-gomez_situ_2023} shares some similarities with our work, the two studies are driven by different motivations. The overlap between the two lies in the classification of \textit{in-situ} vs. \textit{ex-situ} GCs specifically for the MW. Our work has better performance in this area because our algorithms are assessed on a carefully selected sample of galaxies that closely match the MW, ensuring that our results are specifically optimized to the characteristics of the MW. Moreover, we incorporate more kinematic characteristics of GCs derived from the 6D phase space. These inputs prove to be important in identifying \textit{in-situ} and \textit{ex-situ} GCs.

\section{Summary}
\label{sec:summary}

In this work, we evaluate five unsupervised clustering methods, two dimensionality reduction methods, and three supervised classification methods on their ability to classify the progenitors of GCs in the MW-analog systems. We perform the test on four simulated galaxies carefully selected to match the mass assembly history of the MW (Fig.~\ref{fig:mh_log_vs_tlookback}). Two of them are TNG50 galaxies, and the other two are from the Latte suite of the FIRE-2 simulations. We generate realistic GC systems for these galaxies using a cluster formation and evolution model by \citetalias{chen_catalogue_2024} (\S\ref{sec:model_sample}). The model outputs 10 important properties of GCs, including spatial, kinematic, and age information, that are all observable in the MW system.

We begin with dividing model GCs into \textit{in-situ} and \textit{ex-situ} groups (\S\ref{sec:in_vs_ex}). For this purpose, we find that $r,\E,\feh,\varepsilon,$ and $\Jp=L_z$ are the most important variables (Fig.~\ref{fig:feh_logJr}), while $\Jz$ and $\rperi$ are not of particular significance. For the five clustering methods, Agglomerative Clustering and BIRCH show best performance, whereas GMM and Spectral Clustering cannot yield acceptable classification results for all four sample galaxies. The \curse\ becomes notable for the direct clustering methods with more than six input variables (Fig.~\ref{fig:dim_acc}), where adding more data does not improve the classification accuracy. We address this problem by introducing an additional dimensionality reduction step before clustering (Fig.~\ref{fig:dim_acc_reduce}). However, the dimensionality reduction step does not further improve the performance of direct clustering. We also examine a hybrid approach incorporating unsupervised and supervised methods. Such an approach demonstrates no improvement over the other approaches.

Next, we concentrate on the \textit{ex-situ} clusters to distinguish GCs from different progenitor satellite galaxies. To identify the last major merger (GS/E-analog), we cluster the \textit{ex-situ} population into two groups representing the GS/E-analog and the other GCs (\S\ref{sec:n_g_2}). This task can be well handled (accuracy $\gtrsim 90\%$) using seven GC properties: $\age,\Jz,\feh,\Jp,\rapo,\varepsilon,$ and $\E$ (\textit{top row} of Fig.~\ref{fig:more_groups}). Compared to the case of identifying \textit{in-situ} and \textit{ex-situ} GCs, two more inputs are needed to distinguish the subtly different distributions of the two groups in the multi-dimensional space. In this case, Agglomerative Clustering and BIRCH are still the most successful clustering methods.

We proceed to identify the second most dominant merger by clustering the \textit{ex-situ} population into three groups, standing for the GS/E-analog, the second most dominant merger, and the other GCs (\S\ref{sec:n_g_greater_than_3}). However, no variable configuration yields a minimum accuracy $>70\%$ by directly splitting the \textit{ex-situ} GCs into three groups (\textit{middle row} of Fig.~\ref{fig:more_groups}). If we assume that the GS/E-analog can be perfectly labeled and sub-divide only the remaining \textit{ex-situ} GCs into two groups, we can achieve higher accuracy with a minimum $>70\%$ (\textit{bottom row} of Fig.~\ref{fig:more_groups}). Separating the \textit{ex-situ} population into even more groups is unfeasible. By computing the BIC utilizing a cross-validation technique, we quantitatively show that the overfitting problem becomes severe with more groups (Fig.~\ref{fig:gmm_ic_vc}).

After evaluating the accuracy of different algorithms on various configurations of GC properties, we apply our results to the Galactic GCs via a Bayesian approach (\S\ref{sec:mw}). We robustly assign 94 of 150 MW GCs with measured metallicities and kinematics to the \textit{in-situ} group. This number agrees closely with the result of \citet{belokurov_-situ_2024}, who effectively merged the ``low-energy group'' identified by \citet{massari_origin_2019} to the \textit{in-situ} population. Most of the information required for this classification is contained in the $E-L_z$ space. We then identify the GS/E merger from the \textit{ex-situ} group, taking into account prior information summarized by \citet{massari_origin_2019}. This results in $\approx30$ GCs associated to the GS/E. We assign only two clusters from the ``high-energy group'' to the GS/E, and classify the rest of them are as \textit{ex-situ} GCs from other mergers, consistent with the claim by \citet{massari_origin_2019}. Finally, we focus on the remaining GCs and identify a relatively clustered group which is closely linked to the Sagittarius dwarf galaxy. We present the final classification results in Fig.~\ref{fig:classification_mw} and Tab.~\ref{tab:apply_to_mw}.

We also investigate the evolution of the distributions of GCs in the IoM space and action space (\S\ref{sec:evolution} and Fig.~\ref{fig:evolve}). GCs from the \textit{early-accreted} merger more than 11.4~Gyr ago tend to have indistinguishable properties from the true \textit{in-situ} ones, because the central galaxy was not massive enough to prevent the similarly massive satellite from penetrating the inner region and heating up the kinematics of \textit{in-situ} GCs. We also find the GS/E-like merger was disruptive enough to perturb the kinematics of the previously accreted GCs. These GCs used to form distinct cores in the IoM and action spaces, but got less distinguishable after the dominant merger. After the last major merger, all the kinematic variables remain approximately conserved.

To uncover possible connections between the properties of the accreted satellites and their surviving GC populations, we compute the Pearson correlation coefficients between pairs of important parameters (\S\ref{sec:merger_properties} and Fig.~\ref{fig:cov_map_merger_vs_gc}). We find that the maximum mass of the satellite progenitor at infall follows a linear relation with the total mass of its surviving GCs (Fig.~\ref{fig:Mgcs_Mh}). Remarkably, this relation is the same as the previously known $\Mgc$--$\Mh$ relation for the surviving galaxies at $z=0$  \citep{harris_dark_2015,forbes_extending_2018}. We also find that the merger mass ratio correlates with the energy dispersion of GCs, the infall time correlates with the mean age and anti-correlates with the age dispersion, the dissolution time of the satellite anti-correlates with the mean of $\logr$ and $\logJr$, and the merger duration correlates with these two parameters.

We compare our work with \citet{trujillo-gomez_situ_2023}, who developed a supervised classifier for \textit{in-situ} vs. \textit{ex-situ} clusters using the E-MOSAICS simulation (\S\ref{sec:T23}). Our work outperforms their accuracy in the area of identifying \textit{in-situ} and \textit{ex-situ} GCs for MW-analogs because we evaluate our algorithms on the galaxy sample carefully selected to match the MW, and we take into account more kinematic variables that are measurable for the MW GCs.

Finally, we make the source code, raw data, and key results available in public repositories. The source code for the GC formation model can be accessed at \url{https://github.com/ybillchen/GC_formation_model}. The model catalog of GCs is published at \url{https://github.com/ognedin/gc_model_mw}. Additionally, the ASCII version of Tab.~\ref{tab:apply_to_mw} summarizing our classification of MW GCs is accessible at \url{https://umich.edu/~ognedin/mw_gc_classification.txt}.

\begin{acknowledgements}
\small
We are grateful to Andrew Wetzel and Pratik Gandhi for providing the merger trees of the FIRE-2 simulations. We thank Eric Bell, Leandro Beraldo e Silva, Monica Valluri, Andrey Kravtsov, and Yuan-Sen Ting for insightful discussions. We thank the reviewer for their suggestions, which have improved the quality of this work. OG and YC were supported in part by the U.S. National Science Foundation through grant AST-1909063 and by National Aeronautics and Space Administration through contract NAS5-26555 for Space Telescope Science Institute program HST-AR-16614.

\paragraph{Software}
\textsc{numpy} \citep{harris_array_2020}, \textsc{matplotlib} \citep{hunter_matplotlib_2007}, \textsc{scipy} \citep{virtanen_scipy_2020}, \textsc{scikit-learn} \citep{pedregosa_scikit-learn_2011}, and \textsc{agama} \citep{vasiliev_agama_2019}.
\end{acknowledgements}

\bibliographystyle{aasjournal}
\bibliography{GC-model-references}

\appendix

\section{Minimum achievable accuracy}
\label{sec:accuracy}

In this section, we present a rigorous definition of the classification accuracy and derive a mathematical lower limit of accuracy. Considering the case of classifying the data set into $N_{\rm g}$ groups, we name them the $1,2,\cdots,N_{\rm g}$-th classified group. By construction, the number of true groups equals $N_{\rm g}$: in the case of \textit{in-situ} vs. \textit{ex-situ} classification, there are 2 classified groups and 2 true groups; in the case of classifying progenitors of \textit{ex-situ} clusters, we label $N_{\rm g}-1$ true mergers and combine the remaining as an additional group standing for the other mergers. We label the true groups as the $1,2,\cdots,N_{\rm g}$-th true group. We count the number of data points in $i$-th true group being assigned in the $j$-th classified group as $N_{ij}$, forming an $N_{\rm g}\times N_{\rm g}$ matrix. For clarity, we always use the Latin symbol ``$i$'' to refer to the row index and ``$j$'' for the column index.

Next, we aim at finding a mapping from the classified groups to the true groups. We require the mapping to be both one-to-one and onto. This means that every classified group maps to a unique true group; and every true group inversely maps to a unique classified group. All mappings form a permutation group $S_{N_{\rm g}}$.

To select the mapping that best describes the classification results, we search for the mapping $\sigma$ that maximizes the number of correctly classified data points, which is defined as
\begin{equation*}
    N_\sigma\equiv\sum_{j=1}^{N_{\rm g}}\left.N_{ij}\right|_{i=\sigma(j)}.
\end{equation*}
We refer to the best mapping as $\sigma_0$. We define the accuracy of the classification as the number of correctly classified data using $\sigma_0$ divided by the total number of data points, 
\begin{equation*}
    {\rm accuracy}\equiv\frac{N_{\sigma_0}}{\sum_{i=1}^{N_{\rm g}}\sum_{j=1}^{N_{\rm g}}N_{ij}}.
\end{equation*}

Now, we derive the lowest reachable value for the accuracy. It is convenient to re-index the rows of the matrix via $\sigma_0$:
\begin{equation*}
    \sigma_0(1)\rightarrow1,\ \sigma_0(2)\rightarrow2,\ \cdots,\ \sigma_0(N_{\rm g})\rightarrow N_{\rm g}.
\end{equation*}
That is, we pick the $\sigma_0(1)$-th row of $N_{ij}$ as the first row of the new matrix, the $\sigma_0(2)$-th row as the second row of the new matrix, and so on until the $\sigma_0(N_{\rm g})$-th old row as the $N_{\rm g}$-th row of the new matrix. This yields a new matrix $N'_{lj}$, with rows permuted from $N_{ij}$. Here we use ``$l$'' as the row index for the new matrix; we keep ``$j$'' as the column index because the columns are preserved after the re-indexing. We emphasize that the re-indexing just gives each true group a ``new name'', instead of changing the classification results. By construction, we have 
\begin{equation*}
    N'_{lj} = \left.N_{ij}\right|_{i=\sigma_0(l)} \text{ and } N_{ij}=\left.N'_{lj}\right|_{l=\sigma_0^{-1}(i)}.
\end{equation*}
Based on this, we can easily derive that
\begin{equation}
    N'_{\sigma'}=N_\sigma
    \label{eq:conserve}
\end{equation}
where $\sigma'\equiv \sigma_0^{-1}\cdot \sigma$ (for example, $\sigma'_0\equiv \sigma_0^{-1}\cdot \sigma_0=I$). These relations are valid for all indices $l$ and mappings $\sigma$. Therefore, the accuracy becomes 
\begin{equation}
    {\rm accuracy}
    =\frac{N'_{\sigma'_0}}{\sum_{l=1}^{N_{\rm g}}\sum_{j=1}^{N_{\rm g}}N'_{lj}}.
    \label{eq:accuracy}
\end{equation}
Let us consider a series of mappings $\{\sigma'_k\}$ defined as
\begin{equation*}
    \sigma'_k(j)\equiv \left\{
    \begin{array}{ll}
        j + k, & \text{if}\ j + k \leq N_{\rm g} \\
        j + k - N_{\rm g}, & \text{if}\ j + k > N_{\rm g}
    \end{array}
    \right.
\end{equation*}
That is, $\sigma'_k(j)$ shifts the index $j$ by $k$ with a periodic boundary condition. Note that $\sigma'_0$ is naturally included in this series as the case of $k=0$. We can prove that 
\begin{align*}
    \sum_{k=0}^{N_{\rm g}-1} N'_{\sigma'_k} 
    &\equiv \sum_{k=0}^{N_{\rm g}-1}\sum_{j=1}^{N_{\rm g}} \left.N'_{lj}\right|_{l=\sigma'_k(j)} 
    = \sum_{j=1}^{N_{\rm g}}\sum_{k=0}^{N_{\rm g}-1} \left.N'_{lj}\right|_{l=\sigma'_k(j)} \\
    &= \sum_{j=1}^{N_{\rm g}}\left(\sum_{k=0}^{N_{\rm g}-j} \left.N'_{lj}\right|_{l=j+k} + 
        \sum_{k=N_{\rm g}-j+1}^{N_{\rm g}-1} \left.N'_{lj}\right|_{l=j+k-N_{\rm g}}\right)
    = \sum_{j=1}^{N_{\rm g}}\left(\sum_{l=j}^{N_{\rm g}} N'_{lj} + \sum_{l=1}^{j-1} N'_{lj}\right) \\
    &= \sum_{j=1}^{N_{\rm g}}\sum_{l=1}^{N_{\rm g}}N'_{lj} 
    = \sum_{l=1}^{N_{\rm g}}\sum_{j=1}^{N_{\rm g}}N'_{lj}. 
\end{align*}
The last equality of the equation equals the denominator of Eq.~(\ref{eq:accuracy}), i.e., the total number of data points. Using Eq.~(\ref{eq:conserve}) and the definition of $\sigma_0$, we obtain $N'_{\sigma'_k}\leq N'_{\sigma'_0}$ for all $k$. Plugging this inequality and the equation above into Eq.~(\ref{eq:accuracy}), we get
\begin{equation}
    {\rm accuracy}=\frac{N'_{\sigma'_0}}{\sum_{k=0}^{N_{\rm g}-1} N'_{\sigma'_k}}\geq\frac{N'_{\sigma'_0}}{\sum_{k=0}^{N_{\rm g}-1} N'_{\sigma'_0}}=\frac{1}{N_{\rm g}}.
    \label{sec:inequality}
\end{equation}
So far we have proved that the accuracy is always greater than or equal to $1/N_{\rm g}$. The equality can be reached if all $N'_{S'_k}$ are the same, corresponding to completely random classification. Thus, any classification with accuracy close to this value should be regarded as ineffective.

\section{Maximum achievable accuracy}
\label{sec:max_accuracy}

\begin{figure}
    \centering
    \includegraphics[width=0.5\linewidth]{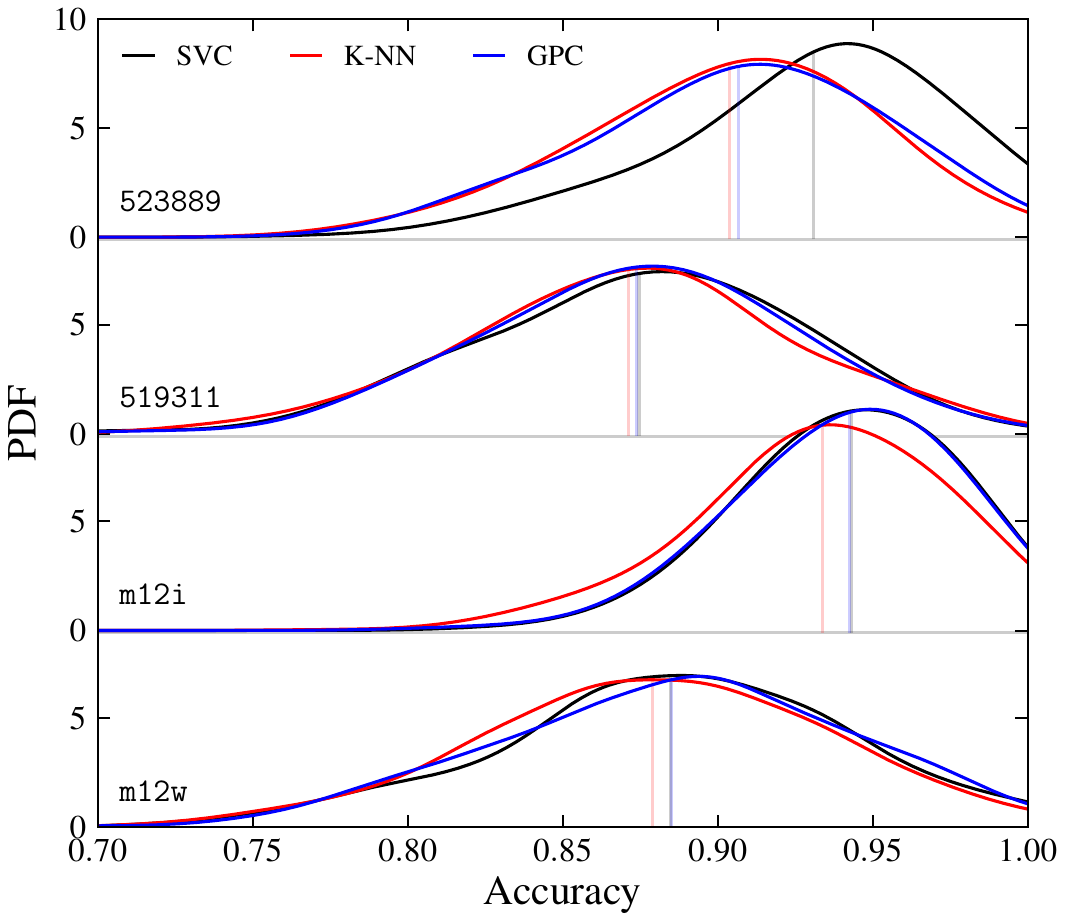}
    \vspace{0mm}
    \caption{Probability density distribution of supervised classification accuracy for the validation sets. We employ a cross-validation with 80\% test set and 20\% validation set, and repeat 500 times to obtain the statistics. The density distribution is approximated via a Gaussian KDE with bandwidth $=0.02$. We calculate the accuracy separately for each sample galaxy, shown in different rows. The vertical lines stand for the mean values of the corresponding PDFs.}
    \label{fig:max_acc}
\end{figure}

Although the classification accuracy by definition (Appendix~\ref{sec:accuracy}) can reach 100\% in ideal cases, achieving a perfect accuracy may not be feasible due to the intrinsic mixing of distributions in the multi-dimensional data space. To address this, we conduct a cross-validation examination utilizing supervised classification methods to determine a more realistic and statistically significant upper bound of accuracy. We take the \textit{in-situ} vs. \textit{ex-situ} case as a case study to derive the maximum achievable accuracy.

To estimate the maximum achievable accuracy, we employ a cross-validation technique similar to the approach in \S\ref{sec:n_g_greater_than_3}. This process splits the data into a training set and a validation set. We train a supervised classifier on the training set and evaluate the classification accuracy on the validation set. We allocate 80\% of the data to the training set and the remaining 20\% to the validation set. This ensures that the training set is large enough for the classifier to learn adequate information. We repeat the training and validation process 500 times to fully sample the accuracy distribution on the validation sets.

In Fig.~\ref{fig:max_acc}, we plot the accuracy distribution on validation sets for our four sample galaxies. We obtain the distribution using three different supervised classification methods detailed in \S\ref{sec:supervised_classification}. The distribution depends weakly on the choices of method, indicating that the limiting factor is indeed the intrinsic mixing of the data, not the specific classification technique. Hence, these accuracy values represent the maximum achievable accuracy.

For the four galaxies in our study, the maximum achievable accuracy ranges from 87\% to 95\%, with a median around 90\%. Any categorization using the same 10 input variables is unlikely to significantly exceed these values without overfitting the data. Our best accuracy values in \S\ref{sec:in_vs_ex} are very close to these upper bounds, suggesting that we are approaching the theoretical maximum with the 10 variables. We also note a galaxy-to-galaxy variation, where \texttt{m12i} exhibits a maximum achievable accuracy greater than 93\%, while that for \texttt{519311} is around 87\%. This agrees with our findings in \S\ref{sec:in_vs_ex}, where \texttt{519311} often shows lower accuracy. This indicates that the extent of phase mixing within the 10D property space differs among galaxies.

\end{document}